\documentclass[a4paper,twoside,prd,nofootinbib,preprintnumbers,twocolumn]{revtex4-1}

\usepackage{amssymb,amsmath,bm,natbib}
\usepackage[table]{xcolor}
\usepackage{slashed}
\usepackage{graphics}
\usepackage{graphicx}
\usepackage{babel}
\usepackage{bbold}
\usepackage[utf8]{inputenc}
\usepackage[caption=false]{subfig}
\usepackage{hyperref}
\usepackage{url}
\usepackage{dsfont}
\usepackage{float} 
\usepackage{cancel}
\usepackage{bbm}
\usepackage[capitalize]{cleveref}
\usepackage{soul}
\usepackage{hhline}

\newcommand{\e}{\ensuremath{\mathrm{e}}}

\begin{document}
\preprint{ZU-TH 02/24, IFT-UAM/CSIC-23-162}
\title{$SU(2)_L$ deconstruction and flavour (non)-universality}

\author{Bernat Capdevila}
\email{bernat.capdevila.soler@gmail.com}
\affiliation{Università di Torino and INFN Sezione di Torino, Via P. Giuria 1, Torino I-10125, Italy,}
\affiliation{DAMTP, University of Cambridge, Wilberforce Road, Cambridge, CB3 0WA, United Kingdom}
\affiliation{Department of Physics and IFAE, Universitat Aut\`onoma de Barcelona, Edifici Ciències, Bellaterra, 08193, Barcelona, Spain}

\author{Andreas Crivellin}
\email{andreas.crivellin@psi.ch}
\affiliation{Paul Scherrer Institut, CH--5232 Villigen PSI, Switzerland}
\affiliation{Physik-Institut, Universit\"at Z\"urich, Winterthurerstrasse 190, CH--8057 Z\"urich, Switzerland}

\author{Javier M. Lizana}
\email{jmlizana@ift.csic.es}
\affiliation{Physik-Institut, Universit\"at Z\"urich, Winterthurerstrasse 190, CH--8057 Z\"urich, Switzerland}
\affiliation{Instituto de F\'isica Te\'orica UAM/CSIC, Nicolas Cabrera 13-15, Madrid 28049, Spain}

\author{Stefan Pokorski}
\email{Stefan.Pokorski@fuw.edu.pl}
\affiliation{Institute of Theoretical Physics, Faculty of Physics, University of Warsaw, Pasteura 5, PL-02-093 Warsaw, Poland}

\begin{abstract}
We study two-site deconstructions of the $SU(2)_L$ gauge group factor of the SM. Models based on this approach can explain the hierarchies of the quark masses and CKM mixing between third and light families if these fields are localised on different sites by the presence of hierarchical new physics scales. The model leads to an accidental global $U(2)_q\times U(3)_u\times U(3)_d$ flavour symmetry which prevents dangerously large effects in flavour observables, making a TeV extension of the SM possible. Given the structure of the PMNS matrix in the neutrino sector, we explore different possibilities for the arrangement of the leptons on the two sites, and consider different models with  $U(2)_{\ell}$ or $U(3)_{\ell}$ flavour symmetries. The phenomenology of the models is mostly governed by a massive vector triplet of $SU(2)_L$. We study the interesting interplay between LHC searches and precision observables. In particular, one of the models can give a sizeable lepton flavour universal effect in the Wilson coefficient $C_9$ while naturally suppressing contributions to $C_{10}$, as suggested by current $b\to s\ell^+\ell^-$ data, predicting simultaneously a mild positive shift in the $W$ boson mass.
\end{abstract}
\maketitle

\newpage
\section{Introduction}

The flavour sector of the Standard Model (SM) seems ad-hoc; quarks and charged lepton masses as well as the CKM matrix elements display a strong hierarchy. Many dynamical explanations have been proposed to explain these features, e.g.~based on adding new horizontal symmetries such as Froggatt-Nielsen models~\cite{Froggatt:1978nt} or gauged flavour symmetries~\cite{King:2003rf,Buras:2011wi}, or new strong dynamics such as anarchic partial compositeness~\cite{Kaplan:1991dc,Grossman:1999ra,Gherghetta:2000qt}. However, stringent bounds from flavour physics, in particular kaon and $D^0-\bar D^0$ mixing or electric dipole moments~\cite{Keren-Zur:2012buf,Panico:2016ull}, typically impose stringent lower limits on the scale of new physics (NP) that realises these mechanisms. 
A very interesting possibility is that the SM flavour structure is generated at different scales~\cite{Berezhiani:1983de,Berezhiani1:1993fg,Barbieri:1994cx,Dvali:2000ha}, that could happen due to strongly coupled sectors developing several condensates~\cite{Panico:2016ull}, or, as we consider here, because the SM gauge group is extended to a non-universal larger group.
In this setup, an approximate global $U(2)$ flavour symmetry minimally broken can emerge, weakening these stringent flavour bounds~\cite{Barbieri:1995uv,Barbieri:2011ci,Crivellin:2011fb,Barbieri:2012uh,Buras:2012sd,Calibbi:2019lvs}.
Then, the lowest scale can be as low as a few TeV, providing interesting phenomenological connections between low-energy precision experiments and LHC searches as well as possible relations with solutions to the Higgs hierarchy problem. For models establishing this relation, see for instance~\cite{Panico:2016ull,Fuentes-Martin:2020bnh,Fuentes-Martin:2022xnb}.

Beyond the SM theories containing a product of identical gauge groups, sometimes referred to as {\it moose} or {\it quiver }structures, have a long history ~\cite{Georgi:1985hf},\cite{Douglas:1996sw}. An additional motivation for them emerged with the idea of deconstructing (latticising) extra spacial dimensions which leads to dual four-dimensional gauge theories~\cite{Arkani-Hamed:2001nha,Hill:2001ps}, referred to as multi-site models. In such models the SM fermions can be assigned to different sites~\cite{Cheng:2001vd}, in analogy with their localisations in the extra dimension, such that the flavour universality is in general broken.

There have been many proposals to address flavour hierarchies following these ideas based on the deconstruction of the SM gauge group or its UV completions: deconstructions of $SU(3)_c$~\cite{Chivukula:2013kw,Crivellin:2015lwa}, including quark-lepton unification of the third family~\cite{Greljo:2018tuh,Crosas:2022quq,Allwicher:2023aql}, theories with a deconstructed Pati-Salam gauge symmetry (totally or partially)~\cite{Bordone:2017bld,Blanke:2018sro,Davighi:2023iks}, deconstructions of hypercharge~\cite{FernandezNavarro:2023rhv,Davighi:2023evx}, $SU(5)$ GUT~\cite{FernandezNavarro:2023hrf} or $U(1)$ SM extensions~\cite{Barbieri:2023qpf}.

On the other hand, it has been suggested that the $U(2)_q$ flavour-symmetry for the light left-handed (LH) quark doublets is sufficient to explain partially the flavour hierarchies of the quark sector~\cite{Greljo:2023bix} as it only allows for third-family-quark Yukawas and forbids light-family Yukawas and light-heavy CKM mixing angles. 
We propose a model that realises this symmetry as an accidental one due to a moose gauge structure: a two-site model for $SU(2)_L$ (i.e.~$SU(2)_1\times SU(2)_2\to SU(2)_L$), with first and second family quarks on one site, and third-family quarks on the other site. While similar extensions of the SM gauge group have been proposed and studied~\cite{Li:1981nk,Muller:1996dj,Malkawi:1996fs,Shu:2006mm,Chiang:2009kb,Hsieh:2010zr,ATLAS:2024tzc}, we put our emphasis on the use of this gauge symmetry to (partially) address flavour hierarchies while having interesting phenomenological consequences at the same time. By charging the Higgs under the same $SU(2)$ group as the third-family quarks, only third-generation quark Yukawa couplings are allowed at the renormalisable level. Heavier NP, well above the TeV scale, can generate the remaining light-quark Yukawa couplings and mixing angles in a suppressed way. This explains (partially) the hierarchy in the quark sector.

However, the flavour pattern in the lepton sector, in particular the PMNS matrix, is not that well addressed by the analogous $U(2)_{\ell}$ symmetry~\cite{Antusch:2023shi}. Therefore, we consider several models with different arrangements for the leptons on the two sites: we compare the model that realises $U(2)_{\ell}$ with models that preserve the full $U(3)_{\ell}$ symmetry, without addressing flavour hierarchies in the lepton sector (whose explanation is relegated to higher scales). This is achieved by situating all leptons in the same site. Possible gauge anomalies can be cancelled by adding new fermionic degrees of freedom.

The breaking of $SU(2)_1\times SU(2)_2\to SU(2)_L$ results in a heavy $SU(2)_L$ triplet vector field containing a $Z^{\prime}$ and $W^{\prime\pm}$ bosons \cite{deBlas:2012qp,Pappadopulo:2014qza}. As we will see, their masses can be as low as a few TeV, even though they have non-universal couplings to LH SM fermions which introduce new sources of flavour breaking. The resulting rich phenomenology includes Flavour Changing Neutral Currents (FCNC) processes such as $b\to s\ell^+\ell^-$ transitions or meson mixing. Notice that these FCNCs are restricted to the LH sector, as within the SM ones, which allows to keep them under control. Still, they already appear at the tree-level, compensating for the mass suppression w.r.t.~the SM contribution. The Higgs, which is charged under one of the $SU(2)$ factors, interacts with the massive triplet generating a mixing between the SM electroweak (EW) gauge bosons and the $Z^{\prime},~W^{\prime}$ that affects the EW precision observables (EWPO), featuring a nice complementarity between flavour, electroweak, and collider physics.

During the completion of the first version of this manuscript, Ref.~\cite{Davighi:2023xqn} appeared on the \texttt{arXiv} where a full deconstruction of $SU(2)_L$ to $SU(2)_1\times SU(2)_2 \times SU(2)_3$, with one factor for each family was proposed. One of the models discussed in our work corresponds to the infrared limit of Ref.~\cite{Davighi:2023xqn}. We have checked that the phenomenological results we obtain for this model are quantitatively similar to the ones of Ref.~\cite{Davighi:2023xqn}.

The details of the proposed models are presented in \cref{sec:2site}. \cref{sec:U2} is devoted to the phenomenology in the $U(2)$ conserving limit, i.e.~without FCNCs. In \cref{sec:NonU2} we include in our study the relevant flavour observables sensitive to the $U(2)$ breaking, with $b\to s$ transitions being the most important ones, and discuss the current status of the $B$ anomalies~\cite{Capdevila:2023yhq} in the context of our models.
\cref{sec:Projections} is dedicated to discussing the prospects of future measurements which can further explore the relevant parameter space.
Finally, we conclude in \cref{sec:Conclusions}.

\section{2-site Model}
\label{sec:2site}

We consider the gauge group $SU(3)_c\times SU(2)_1\times SU(2)_2\times U(1)_Y$, where $SU(2)_1\times SU(2)_2$ is broken at the (multi-)TeV scale to $SU(2)_L$. The gauge couplings correspondng to  $SU(2)_1$ and $SU(2)_2$ are $g_1$ and $g_2$ respectively. For the breaking $SU(2)_1\times SU(2)_2 \to SU(2)_L$ we introduce the link field $\Phi_{ij}$, a doublet under $SU(2)_1$ and $SU(2)_2$, i.e.~$\Phi \to U_1 \Phi U_2^{\dagger}$, with $U_{1,2}$ being rotations in $SU(2)_{1,2}$ space. This link field could be an elementary scalar, but also a condensate originating from strong dynamics of a composite sector \cite{Fuentes-Martin:2020bnh}. In any case, if it develops a vacuum expectation value (VEV), $\Phi_{ij} = \Lambda \delta_{ij}$, it generates a gauge boson mass matrix, which in the $(W_{\mu}^1,W_{\mu}^2)$ basis is given by
\begin{equation}
\mathbbm{M}^2_W=\frac{\Lambda^2 }{4}
\begin{pmatrix}
g_1^2 & -g_1g_2 \\
-g_1g_2 & g_2^2 
\end{pmatrix}.
\end{equation}
The massless eigenstate corresponds to the SM gauge boson $W_{\mu}^{(0)}$, with a universal coupling
\begin{equation}
  g_L=\frac{g_1 g_2}{\sqrt{g_1^2+g_2^2}}, 
\end{equation}
to SM $SU(2)_L$ doublets, independently if they are doublets of $SU(2)_1$ or $SU(2)_2$ prior to the breaking.\footnote{To get fermions which are doublets of $SU(2)_L$, they must be charged under only one of the two $SU(2)_i$ before the breaking.} In addition there is a massive $SU(2)_L$ triplet $W^{\prime(0)}_{\mu}\sim({\bf 1},{\bf 3})_0$, with squared mass $M_{W^{\prime}}^2=\frac{\Lambda^2}{4}(g_1^2+g_2^2)$ and couplings
\begin{align}
g^\prime_{1}=-\sqrt {{g_1^2} - g_L^2} ,~~~~~~g_{2}^{\prime}=\dfrac{{g_L^2}}{{\sqrt {{g_1^2} - g_L^2} }},
\end{align}
where $g^{\prime}_{1(2)}$ is the coupling to fields located in the first (second) site prior to the breaking. As a convention, we fix the sign of the coupling of $g^{\prime}_{2}$ to be positive, which fixes $g^{\prime}_{1}$ to be negative.
Interestingly, $g^{\prime}_1 g^{\prime}_2=-g_L^2$, making impossible to decouple the massive triplet simultaneously from both sites.

The interactions between the massive vector triplet and the SM fields can be parametrised by the Lagrangian
\begin{align}
\mathcal{L}\supset -&\frac{1}{2}\bigg[ 
g^{q}\sum_{i=1,2} \bar q_L^i \gamma ^\mu  {\sigma_a} q_L^{i}+
g^{q}_{33} \bar q_L^3 \gamma ^\mu  {\sigma_a} q_L^{3}\nonumber\\
& +\sum_{i=1,2,3} g_{ii}^{\ell} \bar \ell_L^i \gamma ^\mu  {\sigma_a} \ell_L^{i} +g^{H} H^{\dagger} \sigma_a i\overleftrightarrow{D}^{\mu} H 
\bigg]\,W_\mu ^{\prime(0) a} ,\label{eq:SimplLag}
\end{align}
where $\sigma_a$ are the Pauli matrices and we have assumed a $U(2)_q$ flavour symmetry for the light LH quarks to evade the stringent flavour bounds in the light families. The different couplings depend on how we arrange the families among the two sites that we discuss next.

\subsection{Quark sector}

To realise the desired $U(2)_q$ symmetry for the light-family quarks, which avoids dangerously large effects in kaon and $D^0-\bar D^0$ physics and, at the same time, explains (part of) the hierarchical structures of the quark masses, we localise the first two quark generations on one site (which we choose to be site 1, the ``light-quark site'') and the third generation on the other site (site 2 or ``top site''). This means that the first two generations are charged under $SU(2)_1$ and the third generation under $SU(2)_2$. This setup endows the model with a $U(2)_q\times U(3)_u \times U(3)_d$ accidental symmetry. If we charge the Higgs doublet under $SU(2)_2$ one can only write third generation Yukawa couplings\footnote{While this model does not address the top-bottom or charm-strange hierarchy (as we discuss below), one could think of UV completions with some approximate (effective) $Z_2$ symmetry under which $d^i_R$ are odd, which suppresses any Yukawa term in the Lagrangian involving RH down quarks.}
\begin{equation}
-\mathcal{L}\supset  y^{(t)}_i \bar q_L^3 H^c u_R^i + y^{(b)}_i \bar q_L^3 H d_R^i.
\end{equation}
Here we can use the freedom to perform rotations between the right-handed (RH) quarks, i.e.~redefining $y^{(t)}_i u_R^i \to y_t t_R$, and $y^{(b)}_i d_R^i \to y_b b_R$. We will thus work in the interaction basis where $u_R^3=t_R$ and $d_R^3=b_R$. Note that these Yukawa couplings have further broken the $U(3)_{u,d}$ flavour symmetries to $U(2)_{u,d}$. This fixes the quark and Higgs couplings of the vector triplet in \cref{eq:SimplLag} to be
\begin{align}
g^q=-\frac{g_L^2}{g_2^\prime },~~~~~g^q_{33}=g^H=g^{\prime}_{2}.\label{eq:gqH}
\end{align}

Higher dimension operators can generate the complete Yukawa couplings of light families and CKM mixing after the $SU(2)_1\times SU(2)_2\to SU(2)_L$ breaking:
\begin{equation}
-\mathcal{L}\supset  \frac{1}{\Lambda^{\prime}} \sum_{\substack{i=1,2 \\j=1,2,3}} \left(
y^{(u)}_{ij} \bar q^i_L \Phi H^c u_R^j+
y^{(d)}_{ij} \bar q^i_L \Phi H d_R^j
\right),\label{eq:lightYuk}
\end{equation}
where $\Lambda^{\prime}$ is some NP scale above $\Lambda$. After breaking $SU(2)_1\times SU(2)_2$ to the SM gauge symmetry, these terms generate Yukawa couplings suppressed by the ratio $\Lambda/\Lambda^{\prime}$. 
The flavour hierarchies between third and light-family quarks, therefore, emerge due to the existence of separated scales with $\Lambda \ll \Lambda^{\prime}$. 
Dynamical explanations of flavour hierarchies between first and second families can be postponed to scales $\sim \Lambda^{\prime}$ and above.
The largest breaking of $U(2)_q$ corresponds to the CKM mixing angle between second and third-generation quarks, i.e.~$V_{cb,ts}$. Assuming $y^{(u)}_{23}\sim y_t$  and $ y^{(d)}_{23}\sim y_b$, this implies that a first layer of NP beyond our model should appear at $\Lambda^{\prime}\sim \Lambda/V_{cb}$.

Several UV completions could generate these effective operators:
\begin{itemize}
\item[(1)] {\bf Extra Higgses:} We may add a scalar doublet of $SU(2)_1$, with hypercharge $1/2$, $H_1$. We can write in the Lagrangian terms like,
\begin{align}
-{\cal L} \supset &\, m_{H_1}^2 |H_1^2| + \mu H_1^{\dagger} \Phi H_{2}\nonumber\\
+& \sum_{\substack{i=1,2 \\j=1,2,3}} \left(
y^{(u)}_{ij} \bar q^i_L H_1^c u_R^j+
y^{(d)}_{ij} \bar q^i_L H_1 d_R^j
\right),
\end{align}
where now $H_2\equiv H$ is the Higgs located in the top-site used before. When $H_1$ is integrated out, the effective Yukawa couplings of \cref{eq:lightYuk} with $\Lambda^{\prime}\sim m_{H_1}^2/\mu$ are generated.\footnote{This UV completion corresponds to a two-Higgs doublet model in the limit of large mass for $H_1$ and small mixing angle~\cite{Branco:2011iw}, so the SM Higgs is mostly the top-site Higgs $H_2$.}
\item[(2)] {\bf Vector-like fermions:} By adding heavy vector-like quarks, $Q_{L,R}$ with the same quantum numbers as $q^3_L$ and mass $m_Q$, we can then write the terms
\begin{align}
-{\cal L} \supset & \sum_{j=1,2,3} \left(
y^{(u)}_{j} \bar Q_L H^c u_R^j+
y^{(d)}_{j} \bar Q_L H d_R^j
\right)\nonumber\\
&+ \sum_{i=1,2}  \lambda_{i}\, \bar q_L^{\,i} \Phi Q_R \,.
\end{align}
Integrating out $Q_{L,R}$ generates the effective Yukawa couplings of \cref{eq:lightYuk} with $\Lambda^{\prime}\sim m_{Q}$.
\end{itemize}
Thus, the SM Yukawa couplings can be written as
\begin{equation}
Y_{u} = 
y_{t}\begin{pmatrix}
\Delta_{u}&& \epsilon_{t} V_q \\
0  && 1 \\
\end{pmatrix} ,~~
Y_{d} = 
y_{b}\begin{pmatrix}
\Delta_{d}&& \epsilon_{b} V_q \\
0  && 1 \\
\end{pmatrix} ,
\end{equation}
where we fix $\epsilon_t-\epsilon_b=1$ and $V_q$ and $\Delta_{u,d}$ are naturally suppressed by $\Lambda/\Lambda^{\prime}$.\footnote{For simplicity we assume that $V_q$ is the same in the up and down sector and that $\epsilon_{t,b}$ are real. This, in fact, is the case if $y_{23}^{u,d}$ are generated by one extra vector-like quark.} It is convenient to write the rotation matrices which transform the interaction basis to the mass eigenbasis, i.e. which diagonalize the up-quark or down-quark Yukawa matrices:
\begin{equation}
Y_{u}=L^{\dagger}_{u} \,\hat Y_{u}\,R_{u},~~~~~Y_{d}=L^{\dagger}_{d} \,\hat Y_{d} \,R_{d},
\end{equation}
where $\hat Y_{u,d}$ are diagonal matrices. Using the freedom of the accidental $U(2)$ flavour symmetries, we choose the interaction basis to be aligned with the down basis for the light LH quarks and identical to the mass basis for the light right-handed (RH) quarks. Then, $V_q=(V_{td}^*,V_{ts}^*)^T$ and the rotation matrices are
\begin{align}
L_u &= 
\begin{pmatrix}
V_{ud} && V_{us} && \epsilon_t V_{ub} \\
V_{cd} && V_{cs} && \epsilon_t V_{cb} \\
\epsilon_t V_{td} && \epsilon_t V_{ts} && 1 \\
\end{pmatrix} + O(V_{us}^4),\\
L_d &= 
\begin{pmatrix}
1 && 0 && -\epsilon_b V_{td}^* \\
0 && 1 && -\epsilon_b V_{ts}^* \\
\epsilon_b V_{td} && \epsilon_b V_{ts} && 1 \\
\end{pmatrix} + O(V_{us}^4),\\
R_u &\approx 
\begin{pmatrix}
1 && 0 && \epsilon_t\frac{m_u}{m_t} V_{ub}  \\
0 && 1 &&  \epsilon_t\frac{m_c}{m_t}V_{cb}  \\
-\epsilon_t\frac{m_u}{m_t} V^*_{ub}  && -\epsilon_t\frac{m_c}{m_t} V^*_{cb}  && 1 \\
\end{pmatrix},\\
R_d &\approx  
\begin{pmatrix}
1 && 0 && -\epsilon_b\frac{m_d}{m_b} V_{td}^* \\
0 && 1 && -\epsilon_b\frac{m_s}{m_b} V_{ts}^* \\
\epsilon_b\frac{m_d}{m_b} V_{td} && \epsilon_b \frac{m_s}{m_b}V_{ts} && 1 \\
\end{pmatrix}.
\end{align}

\subsection{Lepton sector}

To ensure anomaly freedom with minimal particle content, we have to locate two families of leptons in the first site, and the third one in the second site.
This implements an accidental symmetry $U(2)_\ell \times U(3)_e$, which can explain the hierarchy of the  charged lepton masses. Because $y_{\tau}\sim y_b$ and $y_{\mu}\sim y_s$, the natural way to split leptons is choosing $s_1=\{\ell_1,\ell_2\}$ and $s_2=\{\ell_3\}$, where we denote $s_{1(2)}$ the set of lepton fields in site $1$ ($2$) and
the indices of $\ell_i$ are ordered accordingly to the mass of the corresponding charge lepton. The hierarchies between the $\tau$ and light lepton Yukawa couplings is explained in the same way as the hierarchy between third-family and light-family quark Yukawas:
\begin{equation}
-\mathcal{L}\supset  y_{\tau}\bar {\ell}_L^{\,3} H \tau_R +
 \frac{1}{\Lambda^{\prime}} \sum_{\substack{i=1,2 \\j=1,2,3}} \left(
y^{(\ell)}_{ij} \bar {\ell}^i_L \Phi H e_R^j
\right),\label{eq:leptonYuk}
\end{equation}
where we have chosen a basis for $e_R^i$ such that $e_R^3 \equiv \tau_R$ is the only RH charged lepton appearing in the first term. The dimension-5 terms of \cref{eq:leptonYuk} can be generated by the same extra Higgs considered before, or by vector-like leptons with the same quantum numbers as $\ell^3_L$. We will call to this arrangement model 0. Other anomaly free assignments of the leptons to the sites are strongly constrained by lepton flavour universality (LFU) tests between electrons and muons and can generate potentially dangerous $\mu\to e$ flavour violation.

Naive expectations within model 0 however suggest also a hierarchical PMNS matrix, contrary to observations~\cite{Greljo:2023bix, Antusch:2023shi}. Indeed, assuming we do not include more degrees of freedom at the TeV scale, the neutrino sector is described by the effective operators
\begin{align}
-{\cal L} \supset  & \frac{1}{\Lambda_{\nu}} \bigg[
y_{\nu_3}(\bar \ell^3_L H^c) (H^{\dagger} \ell^{3c}_L)\nonumber\\
+&\frac{1}{\Lambda^{\prime}}\sum_{i=1,2} y^{\nu}_{3i} (\bar \ell^3_L   H^c) (H^{\dagger} \Phi^{\dagger}\ell^{i c}_L)\nonumber\\
+& 
\frac{1}{\Lambda^{\prime\,2}}\sum_{i,j=1,2} y^{\nu}_{ij} (\bar \ell^j_L \Phi  H^c) (H^{\dagger} \Phi^{\dagger}\ell^{i c}_L)
\bigg].
\end{align}
Denoting by $m_{\nu_i}$ the masses of the neutrinos, and $\theta_{ij}$ the mixing angle between $i$-th and $j$-th family of neutrinos in the PMNS matrix, this suggests a hierarchy $m_{\nu_2}/m_{\nu_3} \ll \theta_{23}\ll 1$, which is not observed in nature (in particular, $\theta_{ij} \sim O(1)$). Still, it is interesting to consider this model at the TeV because particular UV completions may present mechanisms that undo this hierarchy (see for instance Refs.~
\cite{Fuentes-Martin:2020pww,Greljo:2023bix}). 

However, an orthogonal possibility is to preserve the $U(3)_{\ell}\times U(3)_{e}$ and postpone a dynamical explanation capable to address simultaneously the charged lepton and neutrino hierarchies to a higher scale. For instance, it has been suggested that a $U(2)_e$ symmetry helps to explain the charge-lepton mass hierarchies while allowing large PMNS angles~\cite{Antusch:2023shi}. UV completions realising this symmetry while preserving $U(3)_{\ell}$ would be interesting to explore. Therefore, we will also consider the possibility of locating all leptons on the same site. The resulting gauge anomalies can be cancelled introducing new degrees of freedom, or, in case the breaking to the SM group is triggered by the condensate of a composite sector, via this sector~\cite{Fuentes-Martin:2024fpx}. Here, for concreteness, we introduce heavy leptons $L_{L}$ and $L_{R}$, doublets of one $SU(2)_i$ with hypercharge $-1/2$, which are chiral under $SU(2)_1\times SU(2)_2$ but vector-like under the SM group after breaking of $SU(2)_1\times SU(2)_2\to SU(2)_L$. There are two possible models:
\begin{table*}[t]
\begin{center}
\renewcommand{\arraystretch}{1.4}
\begin{tabular}{|c||c|c||c|c||c|c||c|c|}
\cline{4-9}
\multicolumn{3}{c}{} & \multicolumn{2}{|c||}{Model 0} & \multicolumn{2}{c||}{Model 1} & \multicolumn{2}{c|}{Model 2}\\
\hline
Field & $SU(3)_c$ & $U(1)_Y$ & $SU(2)_1$ & $SU(2)_2$ & $SU(2)_1$ & $SU(2)_2$ & $SU(2)_1$ & $SU(2)_2$\\
\hline
\hline
$q^{1,2}_L$ & $\mathbf{3}$ & $1/6$ & $\mathbf{2}$ & $\mathbf{1}$ & $\mathbf{2}$ & $\mathbf{1}$ & $\mathbf{2}$ & $\mathbf{1}$ \\
$q^3_L$ & $\mathbf{3}$ & $1/6$ & $\mathbf{1}$ & $\mathbf{2}$ & $\mathbf{1}$ & $\mathbf{2}$ & $\mathbf{1}$ & $\mathbf{2}$ \\
$\ell^{1,2}_L$ & $\mathbf{1}$ & $-1/2$ & $\mathbf{2}$ & $\mathbf{1}$ & $\mathbf{2}$ & $\mathbf{1}$ &
$\mathbf{1}$ & $\mathbf{2}$ \\
$\ell^{3}_L$ & $\mathbf{1}$ & $-1/2$& $\mathbf{1}$ & $\mathbf{2}$ & $\mathbf{2}$ & $\mathbf{1}$ &
$\mathbf{1}$ & $\mathbf{2}$ \\
\rowcolor{black!10}[\tabcolsep] $L^r_{L}$ & $\mathbf{1}$ & $-1/2$ & $-$ & $-$ & $\mathbf{1}$ & $\mathbf{2}$ &
$\mathbf{2}$ & $\mathbf{1}$\\
\rowcolor{black!10}[\tabcolsep] $L^r_{R}$ & $\mathbf{1}$ & $-1/2$ & $-$ & $-$ & $\mathbf{2}$ & $\mathbf{1}$ &
$\mathbf{1}$ & $\mathbf{2}$ \\
\hline
\hline
$H$ & $\mathbf{1}$ & $1/2$ & $\mathbf{1}$ & $\mathbf{2}$ & $\mathbf{1}$ & $\mathbf{2}$ &
$\mathbf{1}$ & $\mathbf{2}$\\
\rowcolor{black!10}[\tabcolsep] $\Phi$ & $\mathbf{1}$ & $0$ & $\mathbf{2}$ & $\mathbf{\bar 2}$ & $ \mathbf{2}$ & $\mathbf{\bar 2}$ &
$ \mathbf{2}$ & $\mathbf{\bar 2}$ \\
\hline
\end{tabular}
\end{center}
\caption{Representations of the fermion and scalar fields under the extended gauge group for the three models. The index $r$ runs over the number of will-be vector-like leptons $L_{L,R}$, which is $0$, $1$ and $2$ for models 0, 1 and 2 respectively. 
Beyond-the-SM fields are shown in grey.
We only show the fields charged under $SU(2)_{1,2}$. Fields not shown have the same charge assignments as in the SM.}
\label{tab:Models}
\end{table*}
\begin{itemize}
\item {Model 1}: all SM leptons are in the light-quark site. We have $s_1 = \{\ell^i_L,L_R\}$ and $s_2 = \{L_L\}$. The mass of the vector-like lepton is generated via the term
\begin{equation}
-{\cal L}\supset \lambda_L \bar L_L \Phi^{\dagger} L_R .
\label{eq:YukL1}
\end{equation}
All the lepton Yukawa couplings originate from effective operators, which explains their overall suppression compared to the top quark. Notice that $y_{\tau} \sim y_{c}$.
\item {Model 2}: all SM leptons are in the top site. We have $s_1 = \{L_L^{1,2}\}$ and $s_2 = \{\ell_{L}^a,L_R^{1,2}\}$.  The mass of the vector-like lepton is generated from the term
\begin{equation}
-{\cal L}\supset  (\lambda_L)_{rs} \bar L^r_L \Phi L^s_R .
\label{eq:YukL2}
\end{equation}
All lepton Yukawas are allowed at the renormalisable level in this case.
\end{itemize}
In both cases, we need to assume a suppression of mass terms coupling $\ell_i$ to $L_R$, for which we choose a $Z_2$ parity under which the fields $L_{L,R}$ are odd.

The three models are defined by \cref{tab:Models}, and the couplings of the simplified Lagrangian of \cref{eq:SimplLag} are given by \cref{eq:gqH} together with
\begin{itemize}
\item Model 0: 
\begin{align}
g_{11}^\ell= g_{22}^\ell=-\frac{g_L^2}{g^{\prime}_{2}},~~~~~
g_{33}^\ell= g^{\prime}_{2}.
\label{eq:model0}
\end{align}
\item Model 1: 
\begin{align}
g_{11}^\ell=g_{22}^\ell=g_{33}^\ell=-\frac{g_L^2}{g^{\prime}_{2}},
\label{eq:model1}
\end{align}
\item Model 2: 
\begin{align}
g_{11}^\ell=g_{22}^\ell=g_{33}^\ell=g^{\prime}_{2}.
\label{eq:model2}
\end{align}
\end{itemize}
Integrating out the vector triplet at tree level generates the Wilson coefficients in the SM effective field theory (SMEFT) written in \cref{sec:AppSMEFT}.

In models 1 and 2, for large enough values of the Yukawa couplings $\lambda_L$ of \cref{eq:YukL1,eq:YukL2}, the phenomenology of the triplet vector boson can be studied independently of the vector-like leptons, as we do in the following. We thus leave open the possibility of other UV completions that can fix the gauge anomalies. In any case, for completeness, the phenomenology of the vector-like leptons is briefly discussed in \cref{sec:AppVLL}.

\subsection{$Z-Z^{\prime}$ and $W-W^{\prime}$ mixing}

Once the SM Higgs acquires a VEV, the EW symmetry is broken. The coupling between the massive-vector triplet and the Higgs current in \cref{eq:SimplLag} generates a mass mixing between the SM gauge bosons and the $W^{\prime}$ and $Z^{\prime}$ bosons. The mass eigenstates are thus given by
\begin{align}
W^{\pm}=&\cos \alpha_{WW^{\prime}} W^{(0)\pm} +\sin \alpha_{WW^{\prime}} W^{\prime(0)\pm},\nonumber\\
W^{\prime\pm}=&\cos \alpha_{WW^{\prime}} W^{\prime (0)\pm} -\sin \alpha_{WW^{\prime}} W^{(0)\pm},\\
Z=&\cos \alpha_{ZZ^{\prime}} Z^{(0)} +\sin \alpha_{ZZ^{\prime}} W^{\prime(0)}_3,\nonumber\\
Z^{\prime}=&\cos \alpha_{ZZ^{\prime}} W^{\prime(0)}_3 -\sin \alpha_{ZZ^{\prime}} Z^{(0)} ,\label{ZMixing}
\end{align}
where $Z^{(0)}=c_W W^{(0)}_3+ s_W B$ and the mixing angles are
\begin{align}
\sin \alpha_{WW^{\prime}}=&-\frac{g^H}{g_L}\frac{m_W^2}{M_{W^{\prime}}^2},\\
 \sin \alpha_{ZZ^{\prime}}=&\frac{\sin \alpha_{WW^{\prime}}}{c_W},
\end{align}
where $m_W=g_L v/2$. These mixings will lead to corrections of the couplings of the SM $Z$ and $W$ to fermions, affecting the EWPO. When the triplet is integrated out, these corrections are described in the SMEFT by the Wilson coefficients $C^{(3)}_{Hq}$ and $C^{(3)}_{H\ell}$ (see \cref{sec:AppSMEFT}).

\section{Phenomenology: $U(2)$-preserving observables}
\label{sec:U2}

We first analyse the phenomenology of the three models without taking into account observables related to the breaking of the $U(2)$ flavour symmetries. In particular, $U(2)_q$-breaking observables are sensitive to the misalignment between the interaction and the mass basis, parametrised by $\epsilon_{t,b}$ in the quark sector. Excluding them for now, we can analyse the phenomenology of the other observables as functions of only two parameters: $g^{\prime}_{2}$ and $M_{W^{\prime}}$.

\subsection{LHC constraints}

CMS and ATLAS searched for sequential $Z^\prime$ and $W^{\prime}$ bosons, massive vector bosons with the same couplings as the SM ones. We rescale the branching fractions and production cross sections, taking into account the LHC luminosities that we obtain from \texttt{ManeParse}~\cite{Clark:2016jgm}, to obtain the limits on $M_{W^\prime}$ in our model.
The total widths for the branching fractions are calculated taking into account both fermionic and bosonic decays~\cite{Pappadopulo:2014qza}.\footnote{Here, for the models 1 and 2, we assume that the vector-like leptons $L_{L,R}$ are heavy enough to forbid the decay of the massive vector triplet into them. Notice that if this were not the case, the branching fractions of the triplet to SM fields would decrease, weakening the bounds from direct searches.}
We use $pp\to W^\prime, Z^\prime$ with $W^{\prime}\to \ell \nu$~\cite{ATLAS:2019lsy}, $Z^{\prime}\to \ell^+ \ell^-$~\cite{CMS:2021ctt}, $W^{\prime}\to \tau \nu$~\cite{ATLAS:2021bjk}, $Z^{\prime}\to b\bar b$~\cite{ATLAS:2019fgd}, and $Z^{\prime}\to \tau^+ \tau^-$~\cite{ATLAS:2017eiz} searches, with integrated luminosities of $139-140$\,fb$^{-1}$, except the last one which uses $36$\,fb$^{-1}$. To correct for this difference in luminosities, we do a naive rescaling of $\sqrt{36/140}$ to the cross-section limits to estimate the limit at $140$\,fb$^{-1}$.\footnote{Both signal and background scales linearly in the luminosity $L$. Since statistical relative fluctuations decrease like $1/\sqrt{L}$, the limits on the NP signal cross section typically scale like $ \sqrt{L}$.\label{fn:Scaling}} 
We then show the strongest limit we obtain out of these different channels. Limits on the new bosons in these searches are only given for masses up to $4-7$\,TeV. For heavier masses, we use non-resonant di-lepton and mono-lepton searches to further constrain the parameter space. For this, we implemented a $\chi^2$ as a function of the dimension-6 semileptonic Wilson coefficients (see \cref{sec:AppSMEFT}) using \texttt{HighPT}~\cite{Allwicher:2022mcg}. For large triplet masses, we also include di-jet+photon  searches using $79.8$\,fb$^{-1}$~\cite{ATLAS:2019itm} to constrain four-quark operators of light families, taking the limit of the Wilson coefficient from~\cite{Bartocci:2023nvp},
\begin{equation}
-(5.3\,{\rm TeV})^{-2}< C^{(3)}_{qq}<(4.3\,{\rm TeV})^{-2},~~~(95\%~{\rm C.L.}).\label{eq:Limitsqq3}
\end{equation}

\subsection{Electroweak precision observables}

For the electroweak fit, we employ the \{$\alpha_{EM} , m_Z , G_F $\} input scheme and implement the $\chi^2$ function provided in Ref.~\cite{Breso-Pla:2021qoe} into our analysis. In Ref.~\cite{Breso-Pla:2021qoe}, the $\chi^2$ for the EWPO is given as a function of the parameters $\{\delta g_a\}$, defined in \cref{sec:AppEWFit}, which parametrise the modification of the fermion couplings to EW bosons. In \cref{sec:AppEWFit} we also provide the relation between these parameters and the SMEFT Wilson coefficients. The fit is dominated by the tree-level effect in \cref{sec:AppSMEFT}, but we also include the leading logarithms due to the running from the matching scale $\Lambda$ ($\sim 10\,$TeV) to the EW scale. Regarding the $W$ mass, the PDG average $m^{\rm Exp}_W=(80377\pm 12)\,$MeV~\cite{ParticleDataGroup:2018ovx} and the CDF-II result of $m^{\rm Exp}_W=(80434\pm 9)\,$MeV~\cite{CDF:2022hxs} are not well compatible.\footnote{There is also a new result by ATLAS not included in the PDG average ($80360\pm 16)\,$MeV~\cite{ATLAS:2023fsi}.} Both results imply an enhancement w.r.t.~the SM prediction, $m^{\rm SM}_W=(80361\pm 7)\,$MeV~\cite{Bagnaschi:2022whn}. However, while the first result is compatible within errors, the second one displays a $7\sigma$ tension.
Instead of combining both values, we choose to 
perform the EW fit excluding the $W$ mass as an EW observable, and show instead contours with the predicted change in $m_W$.
We have checked that for models 0 and 1, that predict a negative shift in $m_W$, including the CDF experimental value of $m_W$ can make the EW limit $\sim 2-3\,$TeV stronger in the region dominated by the EW fit.

\subsection{LEP-II $e^+e^- \to \ell^+ \ell^-$ data}
\label{sec:eell}

Four-leptons operators can be constrained by $e^+e^- \to \ell^+ \ell^-$ data of LEP-II obtained with center-of-mass energies above the $Z$-pole. Note that the Wilson coefficients of the Higgs-lepton operators $O_{H\ell}$ also enter in these observables, but are much better constrained by the EW fit. We build our own $\chi^2$ function using the Wilson coefficients of \cref{sec:AppSMEFT} and following the procedure of Ref.~\cite{Falkowski:2015krw} using the formulas of Ref.~\cite{Allanach:2023uxz}, and the data published in Ref.~\cite{ALEPH:2013dgf}. Similar methods have been used in Ref.~\cite{Allwicher:2023shc}.

\subsection{Lepton Flavour Universality Tests}

LFU is only violated in model 0 for $\tau$ versus light leptons. There, the ratios involving $\tau\to \ell\nu\nu$ and $\tau\to K(\pi)\nu$ of \cref{eq:tauenunu,eq:taumununu2,eq:tauKnu2} of \cref{sec:AppLFU} potentially constrain the parameter space. \cref{eq:tauenunu,eq:taumununu2,eq:tauKnu2} show the contribution from the Wilson coefficients of the SMEFT evaluated at the EW scale. In our fit, we include the leading log running from the matching to the EW scale, which is small because there is no QCD running, and the potential contribution from $y_t$ to the running of the relevant Wilson coefficients,
\begin{equation}
\mu\frac{\rm d}{{\rm d}\mu} {[C_{H\ell}^{(3)}]_{ii}}\propto y_t^2(g^q_{33}-g^H),
\end{equation}
cancel out due to the arrangement of the third-family-quark doublet and Higgs field on the sites.

\begin{figure*}[t]
\begin{tabular}{ccc}
Model 0 & Model 1  & Model 2\\
$U(2)_{\ell}$ & (all leptons on the light-quark site)  & (all leptons on the top site)\\ \\
\includegraphics[height=0.33\textwidth]{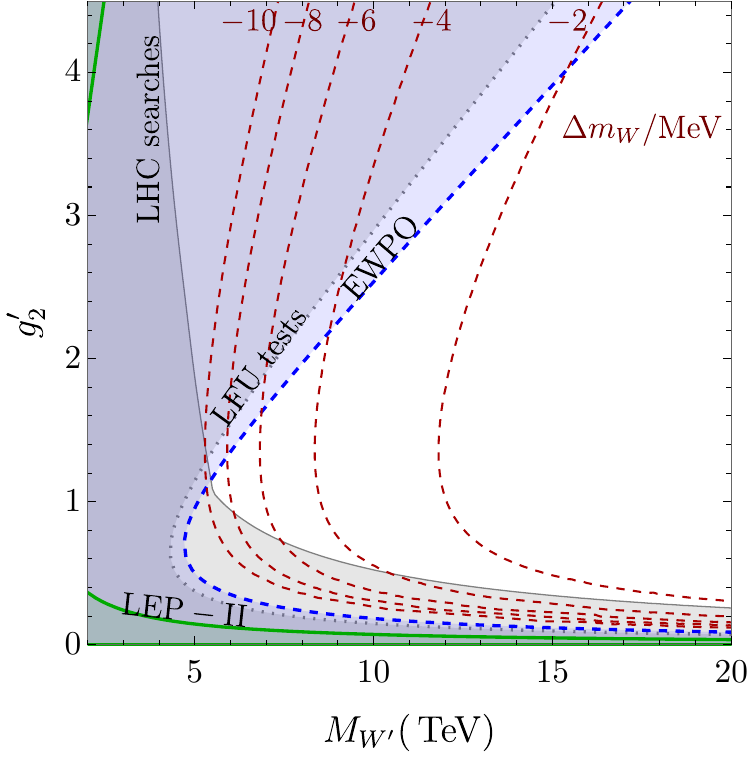} &
\includegraphics[height=0.33\textwidth]{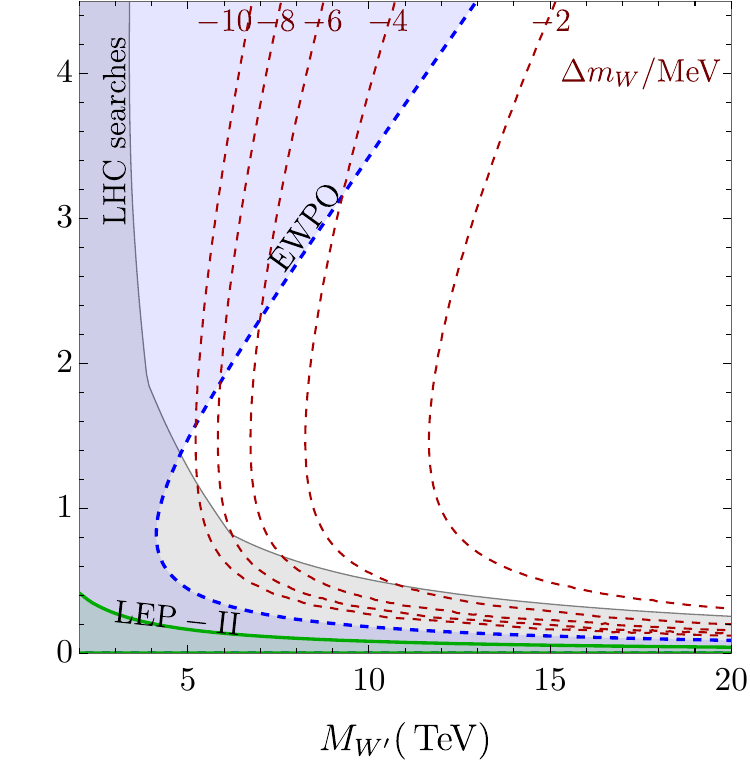} &
\includegraphics[height=0.33\textwidth]{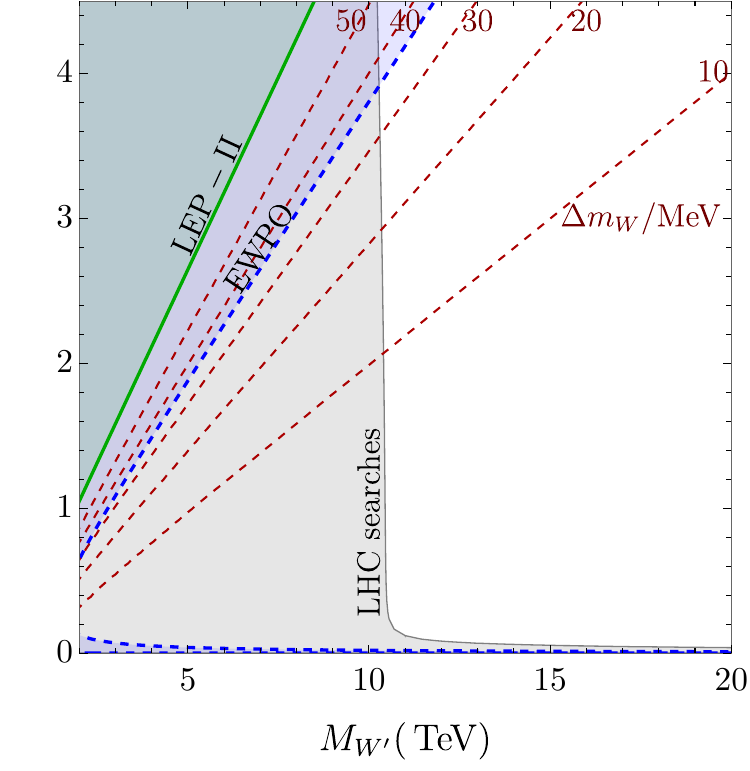}
\end{tabular}
\caption{
Exclusion limits for models 0, 1 and 2 (from left to right) as discussed in \cref{sec:U2}. We include LHC searches (grey solid line), EWPO (blue dashed line) and LEP-II $e^+e^-\to \ell^+ \ell^-$ bounds (green solid line). For model 0 ($\tau$) LFU tests are affected (grey dotted line). Coloured regions are excluded at the 95\% C.L. for 2 d.o.f., except for LFU tests that we use 1 d.o.f. because its $\chi^2$ only depends on one particular combination of the parameters (see \cref{eq:tauenunu,eq:taumununu2,eq:tauKnu2}).
The red dashed contour lines corresponding to $\Delta m_W/{\rm MeV} = (m_W-m_W^{\rm SM})/{\rm MeV}$.}
\label{fig:U2bounds}
\end{figure*}

\subsection{CKM elements}

Our vector triplet, with the couplings given in \cref{eq:SimplLag}, potentially yields NP contributions at the tree level to beta, pion and kaon decays, affecting the extraction of the involved CKM elements~\cite{Crivellin:2020lzu,Capdevila:2020rrl}:
\begin{align}
V^{\rm exp}_{ud}\equiv& \frac{[C^{V,LL}_{\nu e d u}]_{1111}}{[C^{V,LL}_{\nu e}]_{1221}}=\,V_{ud}(1 +\delta V_{ud}), \\
V^{\rm exp}_{us}\equiv& \frac{[C^{V,LL}_{\nu e d u}]_{1121}}{[C^{V,LL}_{\nu e}]_{1221}}=\,V_{us}(1 +\delta V_{us}).
\end{align}
Here $V_{ud}$ and $V_{us}$ are the actual CKM elements within our model, and $C^{V,LL}_{\nu e d u}$, $C^{V,LL}_{\nu e}$ are the Wilson coefficients of the LEFT operators
\begin{align}
O^{V,LL}_{\nu e d u} =&(\bar \nu_L \gamma_{\mu} e_L)(\bar d_L \gamma^{\mu} u_L),\label{eq:VudExp}\\
O^{V,LL}_{\nu e} =&(\bar \nu_L \gamma_{\mu} \nu_L)(\bar e_L \gamma^{\mu} e_L).
\end{align}
Neglecting $V_{ts}$ and $V_{td}$ suppressed contributions, we get
\begin{align}
&\delta V_{ud}
=\delta V_{us}\nonumber\\
&=
v^2\left[[C^{(3)}_{Hq}]_{ii}-
[C_{\ell q}^{(3)}]_{11ii}-[C^{(3)}_{Hl}]_{22}+\frac{1}{2}[C_{\ell\ell}]_{1221}\right]\nonumber\\
&=\frac{(g^H -g^{\ell}_{11}) (g^{\ell}_{22} - g^q)}{g_L^2}\frac{m_W^2}{ M_{W^{\prime}}^2},\label{eq:deltaV}
\end{align}
with $i=1,2$ and we have neglected the small effects from the running of the Wilson coefficients. We can see that for the three models considered here, the various contributions cancel such that $\delta V_{ud}=\delta V_{us}=0$. Possible extensions of the model to address experimental deviations of unitarity in the first row of the CKM matrix, known as the Cabibbo Angle Anomaly (CAA), are discussed in \cref{sec:CAA}.

\subsection{Results}
\label{sec:ResultsIII}

We show the 95\% C.L. exclusion regions of the relevant bounds for our three models in \cref{fig:U2bounds}. For small $g^{\prime}_{2}$, the most relevant bound comes from LHC searches, because a small $g^{\prime}_{2}$ implies large couplings to valence quarks, $g^q=g^{\prime}_{1}=-g_L^2/g^{\prime}_{2}$, and thus large production cross sections for the triplet at LHC. The leading constraint comes from $pp\to \ell \nu$ searches, and, for model 2 and very small $g^{\prime}_{2}$, di-jet searches. It is remarkable that for model 2 most LHC limits are approximately independent of $g^{\prime}_{2}$. The reason for this is that $pp\to W^{\prime} \to \ell \nu$ and $pp\to Z^{\prime} \to \ell^+ \ell^-$ have cross sections $\propto g^q g^{l}_{ii} = -g_L^2$, so that the dependence on the coupling cancels out.

For large $g^{\prime}_{2}$,\footnote{Large values of $g^{\prime}_{2}$ can lead to fast proton decay through non-perturbative effects in this model~\cite{Morrissey:2005uza,Fuentes-Martin:2014fxa}. However, as pointed out in Ref.~\cite{Morrissey:2005uza}, these calculations assume small quartic coupling of the link field $\Phi$. Models with large quartic coupling or realising $\Phi$ as a condensate of a composite sector may avoid this bound.}
the most relevant bound comes from EWPO, although for model 0, also $\tau$-LFU tests are relevant, and for model 2, EWPO only become stronger than LHC searches for very large $g^{\prime}_{2}$. The key to understanding the EW fit in this region of the parameter space is the NP effect on the very constrained modifications of $Z$ couplings to leptons and $b_L$, which at tree level read
\begin{align}
\delta g^{Ze}_{L\,ii}=&\frac{g^Hg^{\ell}_{ii}}{2g_L^2}\frac{m_W^2}{ M_{W^{\prime}}^2} +
\frac{\delta G_F}{4(1-2 s_W^2)},\label{eq:deltageL}\\
\delta g^{Zd}_{L\,33}=&
\frac{ g^H g^q_{33}}{2g_L^2}\frac{m_W^2}{ M_{W^{\prime}}^2}
+\frac{3-4s_W^2}{12(1-2 s_W^2)}
\delta G_F,\label{eq:deltagbL}\\
\delta g^{Ze}_{R\,ii}=&-\delta m_W=
\frac{s^2_W \,\delta G_F}{2(1-2s_W^2)},\label{eq:deltageRmW}
\end{align}
where the precise definition of $\delta x$ is given in \cref{sec:AppEWFit} and
\begin{equation}
\delta G_F =
\frac{g^{\ell}_{11} g^{\ell}_{22} - g^H (g^{\ell}_{11} + g^{\ell}_{22})}{g_L^2}\frac{m_W^2}{M_{W^{\prime}}^2}.
\end{equation}

For model 0, the EW fit is dominated, in the region with sizeable $g^{\prime}_{2}$, by the constraints on the $Z\to\tau_L  \tau_L$ and $Z\to  b_L  b_L$ corrections. This is why within model 1 the limit from EWPO is weaker: the tension in $Z\to \tau_L  \tau_L$ observed in model 0 is removed while the $Z\to  b_L   b_L$ one remains, weakening the limit. Interestingly, contrary to the naive expectation, model 2 further weakens the EW limit. This is because of an accidental approximate cancellation in $Z\to  e^i_L  e^i_L$ between the universal contribution from $\delta G_F$ and the one proportional to $g_{ii}^{\ell}$ occurs, due to the Weinberg angle being close to $\pi/6$, $s_W \approx 1/2$. Also, $Z\to b_L  b_L$ becomes smaller. The leading bound originates from $Z\to e^i_R  e^i_R$, which is affected by the universal contribution from $\delta G_F$. 

Contours of $\Delta m_W = m_W-m_W^{\rm SM}$ show how in models 0 and 1 the $W$ mass is reduced by at most $\sim 10$\,MeV, originating from the shift in $G_F$ (see \cref{eq:deltageRmW}). It is interesting that for model 2, we get an enhancement that can be of about $\Delta m_W\approx 30\,$MeV, going in the direction preferred by the $W$ mass measurements.

\section{Phenomenology: $U(2)$-breaking observables}
\label{sec:NonU2}

Our models necessarily violates quark flavour since $U(2)_q$ is broken (to generate light-heavy CKM elements) through the $V_q$ spurion. This breaking generates FCNCs, in particular, $B_s-\bar B_s$ mixing and $b\to s$ transitions play an important role. Concerning the latter, it is interesting to see if the observed anomalies in $b\to s\ell^+\ell^-$ data and other $B$ decays can be explained in our setup. We also study possible violations of 
$U(2)_{\ell}$ that appear naturally in model 0.

\subsection{$\Delta F=2$ observables}

Meson oscillations are one of the most constraining flavour observables.
Integrating out our vector triplet at tree-level gives contributions to the four-quark operators in LEFT
\begin{align}
\mathcal{L}_{\rm LEFT} \supset & -C_{B_s} (\bar s_L \gamma_{\mu} b_L)^2-
C_{B_d} (\bar d_L \gamma_{\mu} b_L)^2\nonumber\\
& -C_{K} (\bar d_L \gamma_{\mu} s_L)^2
-C_{D} (\bar u_L \gamma_{\mu} c_L)^2.\label{eq:DeltaF2L}
\end{align}
The leading NP contribution to these Wilson coefficients is
\begin{align}
C^{\rm NP}_{B_s}=&V_{tb}^2V_{ts}^{*2}\frac{ (1 - \epsilon_t)^2}{8 M_{W^{\prime}}^2}(g^q_{33} - g^q)^2,\\
C_{B_d}^{\rm NP}=&V_{tb}^2V_{td}^{*2}\frac{ (1 - \epsilon_t)^2}{8 M_{W^{\prime}}^2}(g^q_{33} - g^q)^2,\\
C_{K}^{\rm NP}=&V_{ts}^{2}V_{td}^{*2}\frac{(1 - \epsilon_t)^4}{8 M_{W^{\prime}}^2}(g^{q}_3-g_q)^2,\\
C_{D}^{\rm NP}=&V_{ub}^{2}V_{cb}^{*2}\frac{\epsilon_t^4}{8 M_{W^{\prime}}^2}(g^{q}_3-g_q)^2.\label{eq:DeltaF2Op}
\end{align}
For $B_{d,s}-\bar B_{d,s}$ mixing, assuming that the phases originate only from the CKM elements, we obtain the 95\% C.L.~limits (see \cref{sec:BMixing} for details)
\begin{align}
&-(27.7\,{\rm TeV})^{-2}<\frac{C_{B_s}^{\rm NP}}{ V_{tb}^2V_{ts}^{*2} }<(8.4\,{\rm TeV})^{-2},\label{eq:BsM}\\
&-(13.1\,{\rm TeV})^{-2}<\frac{C_{B_d}^{\rm NP}}{ V_{tb}^2 V_{td}^{*2}}<(7.5\,{\rm TeV})^{-2}.\label{eq:BdM}
\end{align}
For kaon and $D^0-\bar D^0$ mixing, we take the 95\% C.L.~limits from Refs.~\cite{FlavConstraints0,UTfit:2007eik}:
\begin{align}
&-(2.89\times 10^4\,{\rm TeV})^{-2}<{\rm Im} C_{K}^{\rm NP}<(2.04\times 10^4\,{\rm TeV})^{-2},\\
&-(1.03\times 10^4\,{\rm TeV})^{-2}<{\rm Im} C_{D}^{\rm NP}<(1.06\times 10^4\,{\rm TeV})^{-2}.
\end{align}
These observables only depend on the couplings of the new states to the quark sector and are thus the same for the three models, being a test of the main idea behind: a dynamical explanation of the quark hierarchies, without relying on any assumption made for the leptons.

\subsection{$b\to s \ell^+ \ell^+$ data}

Our class of models naturally gives contributions to $b\to s \ell^+ \ell^-$ transitions via the effective operators $O^{\ell}_{9,10}$,
\begin{equation}
\mathcal{L}\supset \frac{2}{v^2}V_{ts}^*V_{tb}\frac{\alpha_{\rm EM}}{4\pi}\sum_{a,i} C^{\ell_i}_{a} O^{\ell_i}_{a},\label{eq:Lagbsll}
\end{equation}
where $a=9,10$, $i=1,2,3$, and
\begin{align}
O^{\ell_i}_{9}=&(\bar s_L \gamma_{\mu} b_L) (\bar \ell_i \gamma^{\mu} \ell_i), \label{eq:O9}\\
O^{\ell_i}_{10}=&(\bar s_L \gamma_{\mu} b_L) (\bar \ell_i \gamma^{\mu} \gamma^5 \ell_i).\label{eq:O10}
\end{align}
These Wilson coefficients do not renormalise under QCD, and the QED running (or the one due to other SM couplings) is negligible. 
The structure of the NP effect is more clearly seen in the basis given by $Z^{(0)}$ and $W^{\prime(0)}_3$, with flavour diagonal couplings. The tree-level exchange of $W^{\prime(0)}_3$ gives contributions proportional to $g_{ii}^\ell$ while the $Z^{(0)}-W^{\prime(0)}_3$ mixing, with $W^{\prime(0)}_3$ coupled to the $bs$ vertex, give universal contributions proportional to $g^H$:
\begin{align}
C_{9}^{\ell_i,{\rm NP}}& = 
-\frac{ \pi  (1-\epsilon_t)}{\alpha_{\rm EM}  }
\frac{(g^{q}_{33}-g^q )[g^{\ell}_{ii}-(1-4s_W^2)g^H] }{g_L^2}
\frac{m_W^2}{M_{W^{\prime}}^2},\label{eq:C9}
\\
C_{10}^{\ell_i,{\rm NP}}& =
\frac{ \pi  (1-\epsilon_t)}{\alpha_{\rm EM}}
\frac{(g^{q}_{33}-g^q )(g^{\ell}_{ii}-g^H) }{g_L^2}
\frac{m_W^2}{M_{W^{\prime}}^2}.\label{eq:c10li}
\end{align}
It is interesting to note that $C_{10}^{\ell_i,{\rm NP}}$ vanishes if $\ell_i$ is located in the same site as the Higgs because the $W_3^{\prime(0)}$ contribution cancels the one from the $Z^{(0)}-W^{\prime(0)}_3$ mixing.\footnote{We have checked that the cancellation of $C_{10}^{\ell_i,{\rm NP}}$ is a general property of $Z^{\prime}$ models when the $Z^{\prime}$ is associated to a $U(1)^{\prime}$ that charges the chiralities of the leptons and the down component of the Higgs such that the $\ell_i$-Yukawa term can be written at the renormalisable level (see \cref{sec:U1protection}). In our case, $U(1)^{\prime}\subset SU(2)_1\times SU(2)_2$ and it is generated by the combination of the generators $(T_3)_1$ and $(T_3)_2$ associated to $W^{\prime(0)}_3$.}

Among the $b\to s\ell^+\ell^-$ observables, $B_s\to \mu^+ \mu^-$ is very constraining due to its precise SM prediction~\cite{Bobeth:2013uxa,Beneke:2019slt}. It receives contributions from $C_{10}^{\mu,{\rm NP}}$:
\begin{equation}
\frac{{\cal B}(B_s\to \mu^+ \mu^-)}{{\cal B}(B_s\to \mu^+ \mu^-)_{\rm SM}}= \left|1+\frac{C_{10}^{\mu,{\rm NP}}}{C_{10}^{\mu,{\rm SM}}}\right|^2,\label{eq:Bsmumu}
\end{equation}
where $C_{10}^{\mu,{\rm SM}}=4.188$ and we take~\cite{Neshatpour:2022pvg,Altmannshofer:2021qrr}
\begin{align}
{\cal B}(B_s\to \mu^+ \mu^-)_{\rm Exp}=&\left( 3.52^{+0.32}_{-0.30} \right) \times 10^{-9},\\
{\cal B}(B_s\to \mu^+ \mu^-)_{\rm SM}=&(3.67\pm 0.15)\times 10^{-9}.\label{eq:BsmumuSM}
\end{align}

After the latest measurement of $R_{K^{(*)}}$ by LHCb~\cite{LHCb:2022vje,LHCb:2022qnv}, $b\to s \ell^+ \ell^-$ data 
do not show indications of lepton flavour universality violation between electrons and muons and are thus compatible with $C_{9e}^{\rm NP}=C_{9\mu}^{\rm NP}\equiv C_9^{\rm U}$ and $C_{10e}^{\rm NP}=C_{10\mu}^{\rm NP}\equiv C_{10}^{\rm U}$. A global fit of $b\to s \ell^+ \ell^-$ data within the NP scenario $(C_9^{\rm U},C_{10}^{\rm U})$ yields~\cite{Alguero:2023jeh,Capdevila:2023hiv}
\begin{equation}
C_9^{\rm U}=-1.18^{+0.18}_{-0.17},~~
C_{10}^{\rm U}=0.10^{+0.13}_{-0.14},
\end{equation}
where $C_{10}^{\rm U}$ is mainly constrained by $B_s\to \mu^+ \mu^-$, which we discuss above, and $C_{9}^{\rm U}$ shows a sizeable negative shift of $\sim 25\%$ w.r.t. its SM prediction. This effect primarily arises from persistent tensions observed in the angular distribution of $B^{(0,+)}\to K^{*(0,+)}\mu^+\mu^-$~\cite{Alguero:2023jeh}, with particular emphasis in two anomalous bins of the so-called $P_5^{\prime}$ observable~\cite{Descotes-Genon:2012isb}. Additionally, we have recently seen the exacerbation of the deviations in the branching ratio of $B^{(0,+)}\to K^{(0,+)}\mu^+\mu^-$ channels, as a consequence of the increased precision in theoretical predictions due to newly available lattice determinations for the relevant form factors across the entire $q^2$ region~\cite{Parrott:2022rgu,Alguero:2023jeh}. Other collaborations involved in $b\to s\ell^+\ell^-$ global fits obtain different results depending on the experimental data input considered and theoretical assumptions regarding the non-perturbative effects that enter into the calculations~\cite{Greljo:2022jac,Ciuchini:2022wbq,Hurth:2023jwr}. However, all analyses that employ available theoretical calculations, based on light-cone sum rules (LCSRs)~\cite{Khodjamirian:2010vf,Gubernari:2020eft}, for the determination of the relevant non-perturbative form factors, coincide in finding a $C_9^\mathrm{U}\sim -1$ and $C_{10}^\mathrm{U}\sim 0$.

\subsection{$B\to K^{(*)}\nu\nu$ and $K\to \pi\nu\nu$}

For the $\Delta F=1$ processes $B\to K^{(*)}\nu\nu$, $K_L\to \pi^0\nu\nu$ and $K^+\to \pi^+\nu\nu$ the relevant effective Lagrangian is
\begin{align}
\mathcal{L}_{\rm LEFT} \supset& -\frac{2}{v^2}V_{ts}^*V_{tb} 
\frac{\alpha_W}{2\pi}\sum_i
C_{sb,i}
(\bar s_L \gamma_{\mu} b_L) (\bar \nu^i_L \gamma^{\mu}\nu^i_L)\nonumber\\
&-
\frac{2}{v^2}V_{td}^*V_{ts} 
\frac{\alpha_W}{2\pi}\sum_i
C_{ds,i}
(\bar d_L \gamma_{\mu} s_L) (\bar \nu^i_L \gamma^{\mu}\nu^i_L).
\label{eq:Lagqqnunu}
\end{align}
with the NP contributions
\begin{align}
C^{\rm NP}_{sb,i}=&\pi^2 (1 - \epsilon_t)\frac{(g^q_{33}-g^q ) (g^H - g^{\ell}_{ii}) }{g_L^2 }\frac{ v^2}{M_{W^{\prime}}^2},\label{eq:Csb}\\
C^{\rm NP}_{ds,i}=&(1-\epsilon_t)\,C^{\rm NP}_{sb,i},\label{eq:Cds}
\end{align}
where the running is negligible. Notice that in our models, $\tilde C_{sb,i}^{\rm NP} =-2 \tilde C_{10}^{\ell_i,{\rm NP}}$, where $\tilde C$ are the Wilson coefficients including their respective normalization factors of~\cref{eq:Lagbsll,eq:Lagqqnunu}.
Likewise, the effect is vanishing if $\ell_i$ is located in the same site as the Higgs. Thus, only models with leptons in the first site (models 0 and 1) will give a contribution.
The resulting branching fractions are~\cite{Buras:1998raa,Buras:2015qea}:
\begin{align}
\frac{{\cal B}( B\to K^{(*)}\nu\bar\nu)}{{\cal B}( B\to K^{(*)}\nu\bar\nu)_{\rm SM}}=&\frac{1}{3}\sum_{i=1}^3
\left|1+\frac{C^{\rm NP}_{sb,i}}{X_t}\right|^2,\\
\frac{{\cal B}( K_L\to \pi^0\nu\bar\nu)}{{\cal B}(  K_L\to \pi^0\nu\bar\nu)_{\rm SM}}=&\frac{1}{3}\sum_{i=1}^3
\left|1+\frac{{\rm Im}(V_{td}^*V_{ts}C^{\rm NP}_{ds,i})}{{\rm Im}(V_{td}^*V_{ts}) X_t}\right|^2,\\
\frac{{\cal B}( K^+\to \pi^+\nu\bar\nu)}{{\cal B}(  K^+\to \pi^+\nu\bar\nu)_{\rm SM}}=&\frac{1}{3}\sum_{i=1}^3
\left|1+\frac{C^{\rm NP}_{ds,i}}{X_t+\frac{V_{us}^4{\rm Re}(V_{cd}^*V_{cs})}{V_{td}^*V_{ts}}P^{(i)}_c}\right|^2,
\end{align}
where $X_t=1.48$, $P^{(1)}_c=P^{(2)}_c=0.45$ and $P^{(3)}_c=0.31$.

We will see that model 0, and to a larger extent model 1, can give an enhancement of up to  $\sim 10\%$ w.r.t.~the SM prediction, which is however still far from the current experimental limit: among the three, ${\cal B}(K^+\to \pi^+ \nu\bar \nu)$ is measured most precisely~\cite{NA62:2021zjw} but still with an error of $\sim 35 \%$, to be increased to $15\%$ by 2025~\cite{Anzivino:2023bhp}. Potential future experiments~\cite{Anzivino:2023bhp} can measure $K^+\to \pi^+ \nu\bar \nu$ (HIKE)~\cite{Ahdida:2023okr} with a $5\%$ error of its SM value, and $K_L\to \pi^0 \nu\bar \nu$ (KOTO-II)~\cite{Aoki:2021cqa} with $20\%$.
Similarly, Belle II aims to measure $B\to K^{(*)}\nu\bar\nu$ within $10\%$~\cite{Belle-II:2018jsg} of its SM value. Note that the current excess observed in $B\to K\nu\nu$ by Belle II~\cite{BKnunu} 
($2.2\,\sigma$), which combined with other measurements suggests an enhancement of about $180\%$ with respect to the SM, is not explainable within our model.\footnote{The differential distributions measured by Belle II are better compatible with the assumption of light NP~\cite{Altmannshofer:2023hkn}.}

\subsection{$R_{D^{(*)}}$}

For completeness, we also discuss
\begin{equation}
R_{D^{(*)}}=\frac{{\cal B}(B\to D^{(*)}\tau\nu)}{{\cal B}(B\to D^{(*)}\ell\nu)},
\end{equation}
with $\ell$ representing light leptons, within our model. The current combined measurements suggests an excess of $10-20\%$ w.r.t~the SM prediction ($\sim$ 3\,$\sigma$)~\cite{HFLAV:2022esi}. Since these observables are a measure of LFU violation, only model 0 can give a contribution. The relevant effective Lagrangian is
\begin{equation}
\mathcal{L}_{\rm LEFT} \supset -\frac{2V_{cb}}{v^2} \sum_{i=1}^3 C_{\ell_i} (\bar \ell_{L}^i \gamma_{\mu} \nu_{L}^i) (\bar c_L \gamma^{\mu} b_L),
\end{equation}
resulting in
\begin{align}
\frac{R_{D^{(*)}}}{R_{D^{(*)}}^{\rm SM}} =&1+2(C_{\tau}^{\rm NP}- C_{e,\mu}^{\rm NP})\nonumber\\
=&1+2\frac{(g^{\ell}_{33} - g^{\ell}_{ii}) (g^q_{33} \epsilon_t + g^q (1 - \epsilon_t)-g^H  )}{g_L^2}\frac{m_W^2}{ M_{W^{\prime}}^2},
\end{align}
with $i=1,2$, and the running is negligible. We observe that the effect in model 0, in the region not excluded by other observables, is below $0.1\%$.

\begin{figure*}[t]
\begin{tabular}{ccc}
Model 0 & Model 1  & Model 2\\
$U(2)_{\ell}$   & (all leptons on the light-quark site)  & (all leptons on the top site)\\ \\
\includegraphics[height=0.33\textwidth]{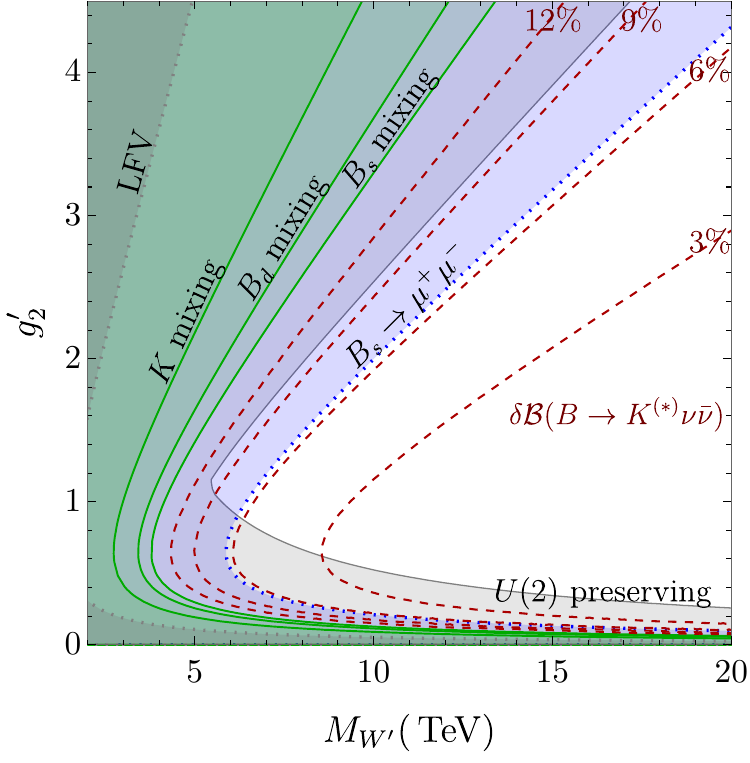} &
\includegraphics[height=0.33\textwidth]{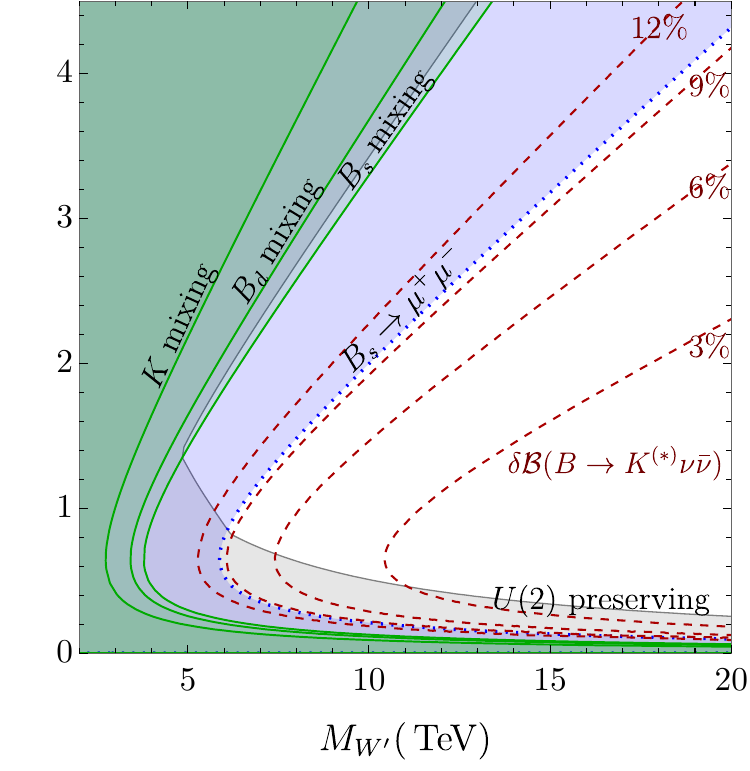} &
\includegraphics[height=0.33\textwidth]{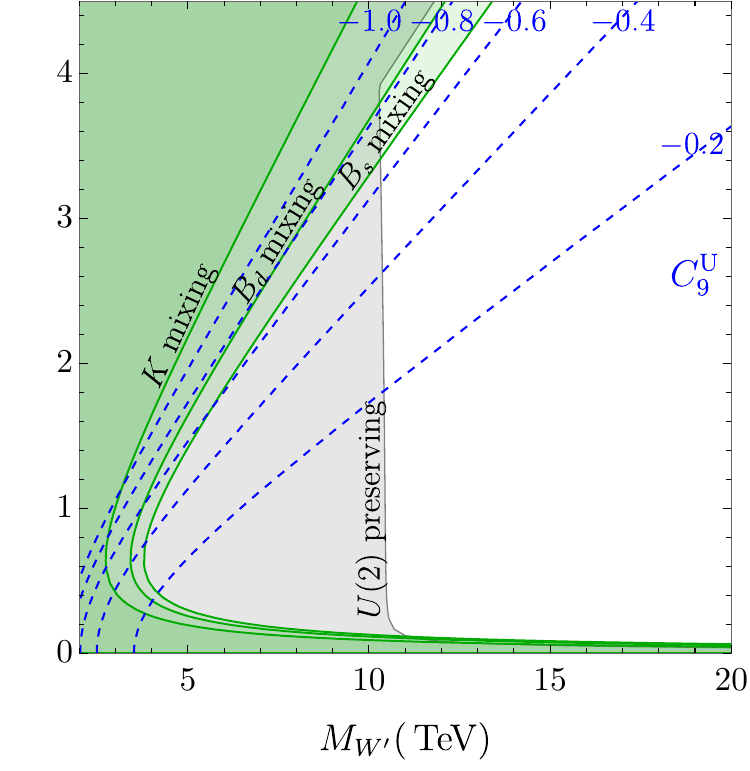}
\end{tabular}
\caption{
Exclusion limits for models 0, 1 and 2 (from left to right) for $\epsilon_t=0$, discussed in \cref{sec:NonU2}. We include the strongest $U(2)$-preserving limit from previous section (grey solid line), $\Delta F=2$ processes (green solid lines), $B_s \to \mu^+ \mu^-$ (blue dotted line), and LFV processes for model 0 assuming \cref{eq:LFVg} is exact (grey dotted line). Coloured regions are excluded at the 95\% C.L. for 2 d.o.f.~except for mesons mixing and $B_{s}\to \mu\mu$ that we take 1 d.o.f.~because their $\chi^2$ only depends on a particular combination of the parameters, $C_{X}^{\rm NP}$ and $C_{10}^{\mu ,{\rm NP}}$ respectively. For model 0 and 1, red dashed contour lines depict $\delta {\cal B}(B\to K^{(*)}\nu\bar\nu)$. The same contour lines give $\delta {\cal B}(K_L\to \pi^0\nu\bar\nu)$ for the same values ($3\%$, $6\%$, $9\%$ and $12\%$) and $\delta {\cal B}(K^+\to \pi^+\nu\bar\nu)$ for an enhancement of $2\%$, $4\%$, $6\%$ and $8\%$ respectively. For model 2, the blue dashed lines show contours of constant value of $C_{9}^{\rm U}$.}
\label{fig:U2BrBounds}
\end{figure*}

\subsection{LFV processes}

Model 0 breaks the $U(3)_{\ell}$ symmetry of the gauge sector of the SM to $U(2)_{\ell}\times U(1)_{\tau}$. Then, it is natural to check contributions to lepton flavour violating (LFV) decays triggered by a possible breaking of $U(2)_{\ell}$.
From \cref{eq:leptonYuk}, we see that, after $SU(2)_1\times SU(2)_2\to SU(2)_L$ breaking, model 0 naturally has the Yukawa couplings
\begin{equation}
\mathcal{L}\supset\frac{\Lambda}{\Lambda^{\prime}}y^{(\ell)}_{i3}\bar\ell^i_L H \tau_R,
\end{equation}
with $i=1,2$. Assuming that $y^{(\ell)}_{i3}\sim y^{(\ell)}_{ii}$, they generate mixing angles between the interaction and the diagonal Yukawa basis, $\theta_{\tau \ell_i} \sim m_{\ell_i}/m_{\tau}$, which induce LFV couplings of the $Z^{\prime}$:
\begin{equation}
g^{\ell}_{i3}\sim \frac{m_{\ell_i}}{m_{\tau}}(g^{\ell}_{33}-g^{\ell}_{ii}),~~g^{\ell}_{12}\sim \frac{m_{\mu}m_e}{m_{\tau}^2}(g^{\ell}_{33}-g^{\ell}_{ii}).\label{eq:LFVg}
\end{equation}
When the triplet is integrated out, these couplings generate the Wilson coefficients of the LEFT operators in \cref{eq:eLeL,eq:eLeR,eq:eLuL,eq:eLdL,eq:eLqR}, giving rise to LFV processes with stringent experimental bounds (see \cref{sec:AppLFV}).
Assuming that~\cref{eq:LFVg} is exact, these bounds are dominated by $\tau^-\to \mu^-\ell^+\ell^-$ in \cref{eq:tau3mu,eq:taumu2e}
and $\mu \to e$ conversion in \cref{eq:muAu2eAu}. We use these observables to build a $\chi^2$-function.

Other LFV decays $\ell_k\to \ell_j \gamma$ receive contributions from the UV matching and from $[C^{(3)}_{H\ell}]_{jk}$ in the SMEFT-LEFT matching, both at one loop level. The first contribution depends on the UV completion realising~\cref{eq:leptonYuk}, which we do not specify here. However, it can be estimated from Appendix A.1 of Ref.~\cite{Capdevila:2020rrl}. We have checked that the bounds from these processes are significantly weaker than the ones discussed above.

\subsection{Results}

All FCNC observables in the quark sector depend crucially on $\epsilon_t$. In the down-alignment limit $\epsilon_t=1$, the effects in the down sector disappear and only $D^0 -\bar D^0$ mixing survives. However, the resulting limit is substantially weaker than the ones from $U(2)$-preserving observables, in particular EWPO. In the following, we will therefore study the limit of alignment in the up-quark sector, $\epsilon_t=0$, resulting in FCNC in the down sector.
In \cref{fig:U2BrBounds} we show the  95\% C.L.~exclusion regions of $B_s\to \mu^+\mu^-$ and meson mixing for the three models, together with the strongest bounds from the $U(2)$-preserving observables discussed in the previous section
and the limits from LFV decays for model 0 assuming~\cref{eq:LFVg} is exact.
We also show the contours lines for $C_9^{\rm U}$ in model 2, and for $\delta {\cal B}(B\to K^{(*)}\nu\bar\nu)$ in models 0 and 1, where 
\begin{equation}
\delta {\cal B}(A)={\cal B}(A)/{\cal B}(A)_{\rm SM}-1.
\end{equation}
The same contour lines also predict the effect in $K_L\to \pi^0\nu\bar\nu$ and $K^+\to \pi^+\nu\bar\nu$ (see caption of \cref{fig:U2BrBounds}). Note that $B\to K^{(*)}\nu\bar\nu$, and $K\to \pi\nu\bar\nu$ is exactly 0 for model 2 (see~\cref{eq:Csb,eq:Cds}). Also, the value of $C_9^{\rm U}$ for models 0 and 1 is strictly positive and at most $\sim 0.1$ in the non-excluded region (see~\cref{eq:C9}).

We see that among $\Delta F=2$ processes, $B_s-\bar B_s$ mixing is the most constraining one, but that all three of them give similar bounds. This is a general property of models with Minimal Flavour Violation (MFV)~\cite{DAmbrosio:2002vsn,EuropeanStrategyforParticlePhysicsPreparatoryGroup:2019qin} or minimally-broken $U(2)$~\cite{Panico:2016ull,Calibbi:2019lvs,Davighi:2023evx,Lizana:2023kei} as ours.
Interestingly, $\Delta F=2$ processes give same order-of-magnitude bounds as the other limits, that, unlike meson mixing, depend on the assumptions made for the leptons in each model.

For the three models the strongest limit for small $g^{\prime}_2$ is again from the $U(2)$ preserving observables (in particular LHC searches) and $\Delta F=2$ processes for very small $g^{\prime}_2$ for model 2. For models 0 and 1, the strongest limit for large $g^{\prime}_2$ is $B_s\to \mu^{+}\mu^{-}$. However, in model 2, $B_s\to\mu^+\mu^-$ remains at its SM value due to the cancellation in the NP contribution to $C_{10}^{\mu}$ observed in \cref{eq:c10li}. This opens the possibility of a large $g^{\prime}_{2}$ coupling for a triplet mass of 10\,TeV-12\,TeV allowing for a sizeable shift in $C_9^{\rm U}$ of about $-0.6$, in the line with the $b\to s \ell^+\ell^-$ fit.
Still, the amount of $C_9^{\rm U}$ is limited by $B_s-\bar B_s$-mixing and the $U(2)$ preserving bounds from EWPO, LHC and LEP-II data. This is similar to what is observed in models with an LFU $Z^{\prime}$ with $bs$ couplings and vector couplings to leptons~\cite{Greljo:2022jac}, which are limited by $B_s$-mixing and LEP-II data. It is interesting to note that in the region where the contribution to $C_9^{\rm U}$ is large we also get an enhancement of the $W$ mass of $\sim 30\,$MeV (see~\cref{fig:U2bounds}).

\section{Future prospects}
\label{sec:Projections}

It is interesting to discuss how future measurements and experiments can explore the parameter space of our three models. For this, we estimate different future limits, assuming that no deviations from the SM prediction are observed. We will comment on the expected sensitivity of different experiments relevant to our models and build the corresponding $\chi^2$ functions to set the projections of the limits in the parameters of the three models in \cref{sec:ProjectionsResults}.

\subsection{LFV: Mu3e, Belle II, COMET and Mu2e}

In the coming years, we expect a significant improvement in limits on LFV processes, affecting the bounds for model 0. Phase-2 of the Mu3e experiment~\cite{Hesketh:2022wgw} aims to reach a sensitivity of $10^{-16}$ at $90\%$ C.L.~in $\mathcal{B}(\mu^+\to e^+e^-e^+)$ in 2028. Belle II will improve the bounds on $\tau$ LFV decays. For an integrated luminosity of $50$ ab$^{-1}$, the projected limits are~\cite{Belle-II:2018jsg}
\begin{align}
\mathcal{B}(\tau^- \to \mu^-\mu^+\mu^-)&<4.6 \times 10^{-10},\\
\mathcal{B}(\tau^- \to \mu^- e^+e^-)&<3.0 \times 10^{-10},
\end{align}
at $90\,\%$ C.L.\footnote{LHCb will also be able to probe branching ratios of $\mathcal{O}(10^{-9})$ of $\tau \to 3\mu$~\cite{LHCb:2018roe}.} The COMET~\cite{COMET:2009qeh} and Mu2e~\cite{Mu2e:2014fns} experiments will test $\mu\to e$ conversion in aluminium with an expected sensitivity of $10^{-16}-10^{-17}$. Conservatively, we take $\mathcal{B}(\mu\,Al \to e \, Al)<10^{-16}$ at $90\%$ C.L.

\subsection{High-luminosity phase of LHC}

We have seen in \cref{fig:U2bounds} that non-resonant searches currently result in the strongest bounds in case of small values of $g_2^{\prime}$. To estimate the limits from the high-luminosity phase of LHC (HL-LHC), we use \texttt{HighPT}~\cite{Allwicher:2022mcg} to extract the projection for the $\chi^2$ for an integrated luminosity of $3\,$ab$^{-1}$ for $pp\to Z^{\prime} \to \ell^+\ell^-$ and $pp\to W^{\prime} \to \ell\nu$. We also include limits on the Wilson coefficient $C_{qq}^{(3)}$  related to the light families from di-jets+photon searches~\cite{Bartocci:2023nvp}. Assuming that the contribution of this operator to the cross section is dominated by its interference with the SM (see Table 1 of Ref.~\cite{Bartocci:2023nvp}), we rescale the limits of~\cref{eq:Limitsqq3} by $\sqrt{79.8/3000}$, as suggested in~\cref{fn:Scaling}, to estimate the HL-LHC bound.

LHCb and CMS will also improve the measurement of $\mathcal{B}(B_s\to \mu\mu)$~\cite{Cerri:2018ypt}. In particular, the expected experimental error of LHCb is $\Delta \mathcal{B}(B_s\to \mu\mu) = 0.16\times 10^{-9}$ for an integrated luminosity of $300\,$fb$^{-1}$~\cite{LHCb:2018roe}. The projected limit is then calculated assuming the theory error in~\cref{eq:BsmumuSM} for the SM prediction.

\begin{figure*}[t]
\begin{tabular}{ccc}
Model 0 & Model 1  & Model 2\\
$U(2)_{\ell}$   & (all leptons on the light-quark site)  & (all leptons on the top site)\\ \\
\includegraphics[height=0.33\textwidth]{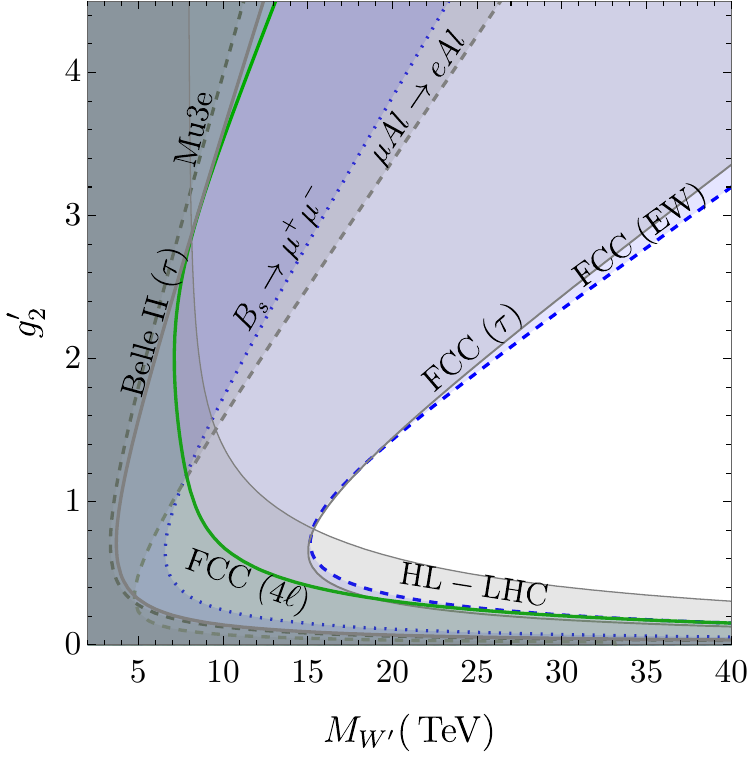} &
\includegraphics[height=0.33\textwidth]{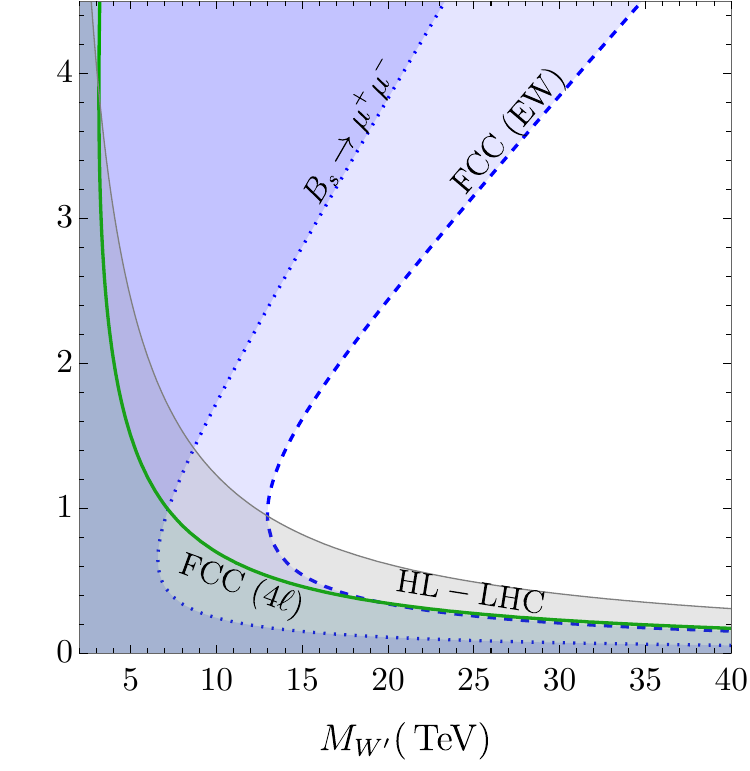} &
\includegraphics[height=0.33\textwidth]{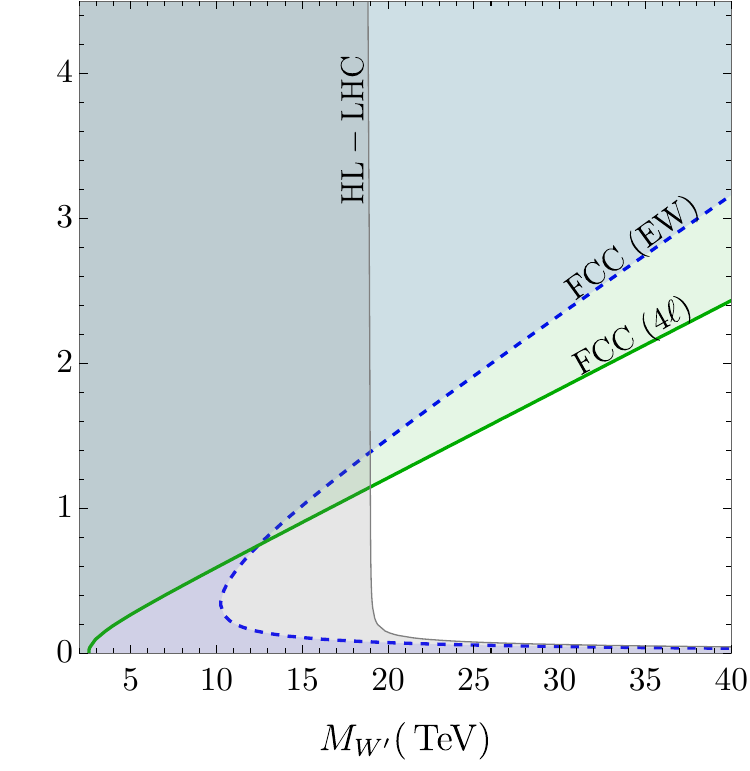}
\end{tabular}
\caption{
Projections of the exclusion limits from different future experiments for models 0, 1 and 2 (from left to right).
We include the LFV limits assuming that \cref{eq:LFVg} is exact in $\mu^+\to e^+e^-e^+$ from Mu3e, $\mu \to e$ conversion from COMET and Mu2e (gray dashed lines) and $\tau$ decays from Belle II (grey solid line), non-resonant searches in HL-LHC (grey solid line), $B_s\to \mu^+ \mu^-$ limit from LHCb assuming $\epsilon_t=0$, and EW observables (blue dashed line), $e^+e^-\to \ell^+ \ell^-$ bounds (green solid line) and $\tau$ LFU tests (grey solid line) from FCC-ee. Coloured regions are excluded at the 95\% C.L.~for 2 d.o.f., except for Mu3e, Belle~II~($\tau$), $\mu\,Al\to e \, Al$, $B_s\to \mu^+ \mu^-$ and $\tau$ LFU tests that we use 1 d.o.f.~because their $\chi^2$ functions only depend on one particular combination of the parameters.}
\label{fig:Projections}
\end{figure*}

\subsection{Future Circular Collider}

The Future Circular $e^+e^-$ Collider (FCC-ee)~\cite{FCC:2018evy,FCC:2018byv} and CEPC~\cite{CEPCStudyGroup:2018ghi} have the potential to improve the limits on the scale of NP coupled to the Higgs boson to several tens of TeV~\cite{Allwicher:2023shc}.
We focus here on the prospects for FCC-ee.

We build a $\chi^2$ with the EW observables of Table 3 of Ref.~\cite{deBlas:2022ofj} ($Z$-pole observables and $W$ boson mass and width) using the \{$\alpha_{EM} , m_Z , G_F $\} input scheme. We add in quadrature the statistical and systematic errors and neglect possible correlations. Table 30 of the same reference also includes the estimated theory uncertainties for most of them. When provided, we also include them, adding in quadrature intrinsic and parametric errors.
Theory uncertainties for the observables $A_{\ell}$ are extracted from $\Delta \sin^2 \theta_W$. We also neglect theory correlations except in the cases of $A_{\ell}$ and $R_{\ell}$, where only one theory error is provided for the three observables associated with the different lepton families, so we assume a full correlation between them. To calculate the contributions of the NP Wilson coefficients to the EW observables, we use the expressions given in Appendix C of Ref.~\cite{Allwicher:2023aql}. 

Table 27 of Ref.~\cite{deBlas:2022ofj} also shows the projected measurements of cross sections and forward-backward asymmetries of $e^-e^+\to \ell^- \ell^+$ at two center-of-mass energies, $s=(240\,$GeV$)^2$ and $s=(365\,$GeV$)^2$. Similarly to \cref{sec:eell}, we use them to build a $\chi^2$ constraining 4-lepton operators.

Finally, FCC-ee would work as a $\tau$ factory and significantly improve the accuracy of $\tau$ decays~\cite{Dam:2018rfz}, relevant for model 0.
FCC-ee can reach a sensitivity of $10^{-10}$ for the LFV $\tau\to 3\mu$ decay. However, regarding $\tau$-physics, the limits for model 0 are dominated by $\tau\to\ell\nu\bar\nu$, with $\ell = e,\mu$, affecting the 
LFU tests (see~\cref{sec:AppLFU}):
\begin{equation}
\Delta R\left[\frac{\tau \to \ell \nu \bar \nu}{\mu \to e \nu \bar \nu} \right] = 
\frac{\Delta \mathcal{B}(\tau\to\ell\nu\bar\nu)}{2 \mathcal{B}(\tau\to\ell\nu\bar\nu)},
\end{equation}
where the expected error is $\Delta \mathcal{B}(\tau\to\ell\nu\bar\nu)=3 \times 10^{-5}$ and we have neglected the error from $\mathcal{B}(\mu\to\e\nu\bar\nu)$.

\subsection{Results}
\label{sec:ProjectionsResults}

In \cref{fig:Projections} we show the $95\%$ C.L.~exclusion limits of the discussed projections for the three models.
Despite the expected improvement in sensitivity on LFV decays from Mu3e and Belle II, we see that for model 0, these bounds will 
still be weaker than
the current $U(2)$-preserving limits of~\cref{fig:U2bounds}.
This is due to the strong suppression of the lepton mixing angles of~\cref{eq:LFVg}.\footnote{Our results for Mu3e are substantially weaker than the ones obtained in Ref.~\cite{Davighi:2023xqn}. The reason is that they assume a CKM-like lepton mixing matrix which gives less suppressed mixing angles for electrons.}
However, $\mu \to e$ conversion will be able to explore the parameter space beyond the current experimental limits. We can also see that, for small $g_2^{\prime}$, non-resonant searches at HL-LHC can improve the limit on the triplet mass by roughly a factor $\sim 2$ for the three models.

FCC-ee will significantly enhance the limits on the triplet mass for moderate and large values of $g_2^{\prime}$. This is especially the case for models 0 and 2, where also $\tau$ LFU tests and $e^+e^-\to \ell^+\ell^-$ observables respectively impose similar or stronger constraints than those coming from the $Z$-pole measurements.

\section{Conclusions}
\label{sec:Conclusions}
A puzzling feature of the quark mass spectrum and the CKM elements is their hierarchy. While the top Yukawa coupling is of order one, all other Yukawa couplings are much smaller. An interesting possibility for understanding this pattern is that only the third-family quark Yukawas are allowed at the renormalizable level while all other couplings are suppressed by higher mass scales, realising a multi-scale explanation of the flavour hierarchies. We implement this idea in a two-site deconstructed model for the $SU(2)_L$ gauge factor, i.e.~$SU(2)_1\times SU(2)_2$ is broken to $SU(2)_L$. The first two generations of quarks are charged under one $SU(2)$ factor while the third family transforms as a doublet under the second factor, together with the Higgs doublet. This only allows third-family quark Yukawas at the tree-level while higher dimensional operators can generate the remaining quark Yukawas in a suppressed way. We discuss different possible UV completions to generate these effective operators, but stay agnostic about the specific realisation. In any case, the model possesses an accidental approximate $U(2)_q$ flavour symmetry which protects FCNC processes from dangerously large effects such that a TeV scale realisation is viable.

The lepton sector differs from the quark sector, in particular, the PMNS matrix is qualitatively different from the CKM matrix. A straightforward extension of the quark structure to the lepton sector predicts LFU violation and would require some particular UV completion to account for the anarchic mixing angles of the PMNS matrix. Another option considered in this paper is to assume that all leptons are charged under the same gauge factor, i.e.~localised on the same site and realising a $U(3)_{\ell}$ symmetry. A dynamical explanation of the lepton hierarchies is then postponed to higher energies. We have then explored the phenomenology of three models with different distributions of the lepton doublets on the sites. 

In the phenomenological analysis, we study the complementary bounds for EWPO, LHC searches and FCNC processes. Interestingly, all those sectors give in general comparable limits on the couplings and masses of the heavy gauge bosons that appear in the models, $W^\prime$ and $Z^\prime$, 
which are in the range of $(5-20)\,$TeV.
We observed several total or partial non-trivial cancellations in NP contributions that weaken or remove some limits, like the limits coming from the determination $V_{ud}$ and $V_{us}$, or the EWPO for model 2. A particularly remarkable cancellation happens in $C_{10}^{\ell_i,{\rm NP}}$, if $\ell_i$ is located on the same site as the Higgs. This allows a sizeable contribution to $C_{9}^{\rm U}$ as preferred by the current $b\to s \ell\ell$ fit, while avoiding the bound from $B_s\to \mu^+\mu^-$. This reveals that, regarding $Z^{\prime}$ bosons, not only those with vector couplings to leptons are the natural candidates to address the deviations observed in $C_9^{\rm U}$~\cite{Greljo:2022jac,Allanach:2023uxz}, but more generally any $Z^{\prime}$ where similar cancellations occur.\footnote{We have checked this is the case for $Z^{\prime}$ bosons with $bs$ couplings and associated to a $U(1)^{\prime}$ that allows for the lepton Yukawas at the renormalisable level (see \cref{sec:U1protection}).}

Possible UV completions of our models to address the hierarchies between first and second families at higher scales could be further deconstructions of the light-family gauge factor~\cite{Davighi:2023xqn}, which in turn, could be UV-completed to models with gauge-flavour unification~\cite{Davighi:2022fer,Davighi:2022bqf}, or realisations of another kind of horizontal gauge symmetries charging the light families and broken in a far UV.

\medskip

\acknowledgments{
The authors would like to thank Marco Ardu, Gino Isidori and Ben Stefanek for useful discussions. The authors also thank Lukas Allwicher for providing codes to implement the LEP-II $e^+e^- \to \ell^+ \ell^-$ data likelihood. 
The work of B.C. is supported by the Margarita Salas postdoctoral
program funded by the European Union-NextGenerationEU. 
The work of A.C. is supported by a Professorship Grant (PP00P2\_176884) of the Swiss National Science Foundation. 
The work of J.M.L. has been supported by the European Research Council (ERC) under the European Union’s Horizon 2020 research and innovation program under grant agreement 833280 (FLAY) in the initial stages of the project, and by the grant CSIC-20223AT023 in the final ones. 
J.M.L. also acknowledges the support of the Spanish Agencia Estatal de Investigacion through the grant “IFT Centro de Excelencia Severo Ochoa CEX2020-001007-S”.
The work of S.P. is  supported
by the National Science Centre, Poland, grant DEC-2019/35/B/ST2/02008.
}

\vspace{5mm}

\appendix

\section{Matching to SMEFT}
\label{sec:AppSMEFT}

We integrate out the heavy triplet at the tree level to match to dimension 6 SMEFT, 
\begin{equation}
\mathcal{L}\supset \sum_n { C}_n O_n,
\end{equation}
and work in the Warsaw basis~\cite{Grzadkowski:2010es}. We have crosschecked our results with \texttt{Matchete}~\cite{Fuentes-Martin:2022jrf}. Neglecting all Yukawa couplings except the top one, and working in the interaction basis, we get at the matching scale the following non-vanishing Wilson coefficients:
\begin{align}
[C^{(3)}_{\ell q}]_{iikk}=&-\frac{g^\ell_{ii} g^q_{kk}}{4M_{W^{\prime(0)}}^2},\\
[C_{\ell \ell}]_{iiii}=&-\frac{(g^{\ell}_{ii})^2}{8M_{W^{\prime(0)}}^2},\\
[C_{\ell \ell}]_{iijj}=&\,\frac{g^{\ell}_{ii}g^{\ell}_{jj}}{4M_{W^{\prime(0)}}^2},\\
[C_{\ell \ell}]_{ijji}=&-\frac{g^{\ell}_{ii}g^{\ell}_{jj}}{2M_{W^{\prime(0)}}^2},\\
[C^{(3)}_{qq}]_{iiii}=&-\frac{(g^{q}_{ii})^2}{8M_{W^{\prime(0)}}^2},\\
[C^{(3)}_{qq}]_{iijj}=&-\frac{g^{q}_{ii} g^{q}_{jj}}{4M_{W^{\prime(0)}}^2},\\
[C^{(3)}_{Hq}]_{ii}=&-\frac{g^{q}_{ii} g^H}{4M_{W^{\prime(0)}}^2},\\
[C^{(3)}_{H\ell}]_{ii}=&-\frac{g^{\ell}_{ii} g^H}{4M_{W^{\prime(0)}}^2},\\
C_{H\Box} =&-\frac{3(g^H)^2}{8M_{W^{\prime(0)}}^2},\\
[C_{uH}]_{33}=&-\frac{y_t (g^H)^2}{4M_{W^{\prime(0)}}^2},\\
C_{H}=&-\frac{\lambda (g^H)^2}{2M_{W^{\prime(0)}}^2},
\end{align}
where $i,k=1,2,3$; $j=2,3$; $i<j$, repeated indices are not summed, and $g^q_{11}=g^q_{22}\equiv g^q$, $g^q_{33}$ and $g^{\ell}_{ii}$ are the couplings of the fermions to the extra triplet depending on the model (see \cref{eq:gqH,eq:model0,eq:model1,eq:model2}).

\section{Vector-like-lepton phenomenology}
\label{sec:AppVLL}

Models 1 and 2 with only the SM fermion fields require of some UV completion to fix their gauge anomalies. One possibility is the inclusion of the vector-like leptons described in \cref{tab:Models}. Up to dimension 5 interactions, for model 1 we can write
\begin{equation}
-\mathcal{L}\supset
\lambda^L \bar L_L \Phi^{\dagger} L_R
+m_i \bar \ell_L^i L_R
+y^L_{i} \bar{L}_L H e_R^i+ {\rm h.c.},  
\end{equation}
and for model 2:
\begin{equation}
-\mathcal{L}\supset
\lambda^L_{rs} \bar L^r_L \Phi L_R^s 
+m_{ir} \ell_L^i L^r_R
+\frac{y_{ri}^{\prime\, L }}{\Lambda^{\prime}} \bar{L}_L^r H\Phi e_R^i + {\rm h.c.},  
\end{equation}
where we assume we work in the basis that diagonalizes $\lambda^L_{rs}=\lambda_r \delta_{rs}$.
It is necessary to assume a ${Z_2}$ parity charging $L_{L,R}$ that suppresses all terms except $\lambda_L$.
After the $SU(2)_1\times SU(2)_2\to SU(2)_L$ breaking, we obtain for both models
\begin{align}
-\mathcal{L}\supset&
M_r\delta_{rs} \bar L^r_L L_R^s+ m_{ir} \bar\ell_L^i L^r_R
+y^L_{ri} \bar{L}_L^r H e_R^i + {\rm h.c.},
\end{align}
where $r,s$ run over the number of vector-like leptons depending on the model, $M_r = \Lambda \lambda_r$, and
$y^L_{ri}= y_{ri}^{\prime\, L}\Lambda/\Lambda^{\prime}$
for model 2.
Although suppressed by the $Z_2$ parity, the parameters $m_{ir}$ and $y^L_{ri}$ can trigger the decay of the vector-like leptons to SM fields through a small mixing with SM leptons.

Direct searches at LHC can put limits on the vector-like-lepton mass through double production. Assuming prompt decay through mixing with the tau, these limits are at the level of the TeV~\cite{CMS:2022nty}, still far from their expected mass $\sim M_{W^{\prime}}$.

Their low-energy impact could be more dangerous, even for small values of $m_{ir}$ and $y^L_{ri}$, mainly because of LFV processes (see \cref{sec:AppLFV}). Here we derive the bounds these couplings need to satisfy to be phenomenologically acceptable.

The mass terms $m_{ir}$ induce non-diagonal couplings of the $Z^{\prime}$ boson,
\begin{equation}
g^{\ell}_{ij} = \sum_r \frac{m_{ri} m^*_{rj}}{M_r^2}\left(g_2^{\prime}+\frac{g_L^2}{g_2^{\prime}}\right).
\end{equation}
They contribute to the LEFT Wilson coefficients given in \cref{eq:eLeL,eq:eLeR,eq:eLuL,eq:eLdL,eq:eLqR} generating LFV processes. In our models, the most constraining observables are $\tau$ LFV decays, $\tau\to \mu \ell \ell$ and $\tau\to e \ell \ell$~(see \cref{eq:tau3mu,eq:tau2mue,eq:taumu2e,eq:tau3e}), and $\mu\to e$ conversion  (see \cref{eq:muAu2eAu}). Varying the value of $g_2^{\prime}\in(g_L^2/3,3)$, we obtain for models 1 and 2 at the $95\%$ C.L.
\begin{align}
\frac{1}{M_{W^{\prime}}^2}\big|\sum_r\frac{m^*_{r1}m_{r2}}{M_r^2} \big|< &\, (0.2-2)\times 10^{-5}\, {\rm TeV}^{-2},\\
\frac{1}{M_{W^{\prime}}^2}\big|\sum_r\frac{m^*_{r2}m_{r3}}{M_r^2} \big|< & \,(0.004-0.1)\, {\rm TeV}^{-2},\\
\frac{1}{M_{W^{\prime}}^2}\big|\sum_r\frac{m^*_{r1}m_{r3}}{M_r^2} \big|< & \,(0.004-0.1)\, {\rm TeV}^{-2}.
\end{align}
Furthermore, integrating out the vector-like leptons, the Yukawa terms $y_L^{ai}$ generate contributions to the SMEFT Wilson coefficients
\begin{equation}
[C_{He}]_{ij}=\sum_{r}\frac{y^*_{ri} y_{rj}}{2M_{r}^2}.
\end{equation}
Flavour-diagonal Wilson coefficients are mostly constrained by EWPO~\cite{Breso-Pla:2021qoe}. For a detailed analysis of the impact of vector-like leptons on the EWPO, see~\cite{Crivellin:2020ebi}. Off-diagonal ones are constrained by the same LFV processes commented above (LFV $\tau$ decays and $\mu \to e$ conversion) through their contribution to the LEFT operators
\begin{align}
[C_{ee}^{V,LR}]_{iijk}=&(2s_W^2-1)[C_{He}]^{jk},\\ 
[C_{ee}^{V,RR}]_{iijk}=&\,2\,s_W^2[C_{He}]_{jk}, \\
[C_{ue}^{V,LR}]_{iijk}=&\left(1-\frac{4}{3}s_W^2 \right) [C_{He}]_{jk}, \\
[C_{eu}^{V,RR}]_{jkii}=&-\frac{4}{3}s_W^2[C_{He}]_{jk},  \\
[C_{de}^{V,LR}]_{iijk}=&\left(\frac{2}{3}s_W^2-1 \right) [C_{He}]_{jk}, \\
[C_{ed}^{V,RR}]_{jkii}=&\frac{2}{3}s_W^2[C_{He}]_{jk},
\end{align}
where $k>j\geq i$.
We thus obtain at the $95\%$ C.L.
\begin{align}
\big|\sum_{r}\frac{y^*_{ri} y_{ri}}{M_{r}^2}\big|<&\, (0.01-0.1)\, {\rm TeV}^{-2},~(i=1,2,3),\\
\big|\sum_{r}\frac{y^*_{r1} y_{r2}}{M_{r}^2}\big|<&\,10^{-5}\, {\rm TeV}^{-2},\\
\big|\sum_{r}\frac{y^*_{r1} y_{r3}}{M_{r}^2}\big|<&\, 0.03\, {\rm TeV}^{-2},\\
\big|\sum_{r}\frac{y^*_{r2} y_{r3}}{M_{r}^2}\big|<&\, 0.02\, {\rm TeV}^{-2}.
\end{align}

\section{Electroweak fit}
\label{sec:AppEWFit}

We work in the \{$\alpha_{EM} , m_Z , G_F $\} input scheme where the relevant terms of the effective Lagrangian for the EW fit are
\begin{align}
\mathcal{L} \supset &- \frac{g_L}{\sqrt{2}}W^{+\mu}\bigg[\bar u_L^i \gamma_{\mu} \left(V_{ij}
+\delta g_{ij}^{Wq}\right)d_L^j 
\nonumber\\
&+ \bar \nu_L^i \gamma_{\mu} \left(\delta_{ij}+\delta g_{ij}^{W \ell }\right)e_L^j\bigg]+{\rm h.c.}\nonumber \\
&-\frac{g_L}{c_W}\, Z^{\mu} \bigg[ \bar f^i_L \gamma_{\mu}
\left( g_L^{Zf}\delta_{ij}+ \delta g_{L\,ij}^{Zf}\right) f_L^j
\nonumber \\
&+\bar f^i_R \gamma_{\mu}
\left( g_R^{Zf} \delta_{ij}+ \delta g_{R\,ij}^{Zf}\right) f_R^j
\bigg] \nonumber \\
&+\frac{g_L^2 v^2}{4}(1+\delta m_W)^2 W^{+\mu} W^-_{\mu}+\frac{g_L^2 v^2}{8 c_W^2}Z^{\mu} Z_{\mu},
\label{EWEffLag}
\end{align}
where
\begin{align}
g_L^{Zf}=\,T_f^3-s_W^2 Q_f ,~~~~~~g_R^{Zf}= -s_W^2 Q_f .
\end{align}
The general relation between the parameters $\delta g$ and $\delta m_W$ with the SMEFT Wilson coefficients in the Warsaw basis can be found in Ref.~\cite{Breso-Pla:2021qoe} and, approximating $V_{\rm CKM}= \mathbb{1}$, is
\begin{align}
\delta g_{L\,ij}^{Z\nu}=&-\frac{v^2}{2}\left([C_{Hl}^{(1)}]_{ij}-[C_{Hl}^{(3)}]_{ij}\right)+\delta^U(1/2,0)\, \delta_{ij},\\
\delta g_{L\,ij}^{Ze}=&-\frac{v^2}{2}\left([C_{Hl}^{(1)}]_{ij}+[C_{Hl}^{(3)}]_{ij}\right)+\delta^U(-1/2,-1)\delta_{ij}\\
\delta g_{R\,ij}^{Ze}=&\,-\frac{v^2}{2}[C_{He}]_{ij}+\delta^U(0,-1)\, \delta_{ij},\\
\delta g_{L\,ij}^{Zu}=&\,-\frac{v^2}{2} \left(
[C_{Hq}^{(1)}]_{ij}-[C_{Hq}^{(3)}]_{ij}
\right)+\delta^U(1/2,2/3)\, \delta_{ij},\\
\delta g_{R\,ij}^{Zu}=&\,-\frac{v^2}{2}[C_{Hu}]_{ij}+\delta^U(0,2/3)\, \delta_{ij}
,\\
\delta g_{L\,ij}^{Zd}=&-\frac{v^2}{2}\left([C_{Hq}^{(1)}]_{ij}+[C_{Hq}^{(3)}]_{ij}\right)+\delta^U(-1/2,-1/3)\, \delta_{ij},\label{eq:ZtobLbL}\\
\delta g_{R\, ij}^{Zd}=&\,-\frac{v^2}{2}[C_{Hd}]_{ij}+\delta^U(0,-1/3)\, \delta_{ij},\\
\delta g^{W\ell}_{ij}=&\,\,\delta g_{L\,ij}^{Z\nu}-\delta g_{L\, ij}^{Ze}\,,\\
\delta g^{Wq}_{ij}
=&\,\,\delta g_{L\,ij}^{Zu}-\delta g_{L\,ij}^{Zd}\,,\\
\delta m_W=&-\frac{v^2 g_L^2}{4(g_L^2-g_Y^2)}C_{HD}
-\frac{v^2 g_L g_Y}{g_L^2-g_Y^2}C_{HWB}\nonumber\\
&-\frac{g_Y^2}{2(g_L^2-g_Y^2)}\delta G_F,
\end{align}
where $\delta^U(T^3,Q)$ is a family-universal contribution that is given by
\begin{align}
\delta^U(T^3,Q)=&
-\left( T^3+Q\frac{g_Y^2}{g_L^2-g_Y^2} \right)
\left(\frac{v^2}{4}C_{HD}+\frac{1}{2} \delta{G_F}\right)
\nonumber\\
&-Q\frac{g_Lg_Y}{g_L^2-g_Y^2}v^2 C_{HWB},
\end{align}
and 
\begin{equation}
\delta{G_F}=v^2\left([C^{(3)}_{H\ell}]_{11}+[C^{(3)}_{H\ell}]_{22}-\frac{1}{2}[C_{\ell\ell}]_{1221}\right)
\end{equation}
is the NP contribution to $G_F$:
\begin{equation}
G_F^{\rm exp}\equiv -\frac{[C^{V,LL}_{\nu e}]_{1221}}{2\sqrt{2}}= \frac{1}{\sqrt{2}v^2}\left(1 + \delta{G_F}\right),
\end{equation}
where $C^{V,LL}_{\nu e}$ is the Wilson coefficient of the LEFT operator $O^{V,LL}_{\nu e}=(\bar \nu_L \gamma_{\mu} \nu_L)(\bar e_L \gamma^{\mu} e_L)$. 

\section{Lepton Flavour Universality Violation in Charge currents}
\label{sec:AppLFU}

In general, extensions of a vector-triplet with couplings like in~\cref{eq:SimplLag} will be constrained by tests of LFU in $\tau$ decays~\cite{HFLAV:2019otj}. 
Defining
\begin{equation}
R\left[ \frac{A}{B}\right] =\left[ \frac{{\cal B}(A)/{\cal B}(A)_{\rm SM}}{{\cal B}(B)/{\cal B}(B)_{\rm SM}}\right]^{\frac{1}{2}},
\end{equation}
we have that
\begin{align}
&{\mkern 1mu} R\left[ {\dfrac{{\tau  \to e\nu \nu }}{{\mu  \to e\nu \nu }}} \right] \nonumber\\
&= 1 + v^2\left([C_{H\ell}^{(3)}]_{33}-[C_{H\ell}^{(3)}]_{22}+\frac{1}{2}([C_{\ell \ell}]_{1221}-[C_{\ell \ell}]_{1331})\right)\nonumber\\
&= 1 + \dfrac{{(g_{11}^\ell-g^H) ( {g_{33}^\ell  - g_{22}^\ell } )}}{{g_L^2}}\dfrac{{m_W^2}}{{{M_{W^{\prime}}^2}}} ,\label{eq:tauenunu}
\end{align}
\begin{align}
&{\mkern 1mu} R\left[ {\dfrac{{\tau  \to \mu \nu \nu }}{{\mu  \to e\nu \nu }}} \right] \nonumber\\
&=1 + v^2\left([C_{H\ell}^{(3)}]_{33}-[C_{H\ell}^{(3)}]_{11}+\frac{1}{2}([C_{\ell \ell}]_{1221}-[C_{\ell \ell}]_{2332})\right)\nonumber\\
&= 1 + \dfrac{(g_{22}^\ell-g^H) ( {g_{33}^\ell  - g_{11}^\ell } )}{g_L^2}\dfrac{{m_W^2}}{{{M_{W^{\prime}}^2}}} ,\label{eq:taumununu2}
\end{align}
\begin{align}
&R\left[ {\dfrac{{\tau  \to K\nu }}{{K \to \mu \nu }}} \right] = R\left[ {\dfrac{{\tau  \to \pi \nu }}{{\pi  \to \mu \nu }}} \right] \nonumber\\
&= 1 + v^2\left([C_{H\ell}^{(3)}]_{33}-[C_{H\ell}^{(3)}]_{22}+[C_{\ell q}^{(3)}]_{22ii}-[C_{\ell q}^{(3)}]_{33ii}\right)\nonumber\\
&= 1 + \dfrac{{(g^q-g^H)( {g_{33}^\ell  - g_{22}^\ell } )}}{{g_L^2}}\dfrac{{m_W^2}}{{{M_{W^{\prime}}^2}}},\label{eq:tauKnu2}
\end{align}
where $i=1$ or $2$. We have included the NP contribution from SMEFT, which should be evaluated at the EW scale, and their value at tree level neglecting running.
\cref{tab:LFUobs} shows the experimental measurements.

\begin{table}[t]
	\begin{tabular}{l c l} \toprule
		Observable & Measurement\\
		\colrule
		$R\left[\frac{\tau\rightarrow e\nu\bar{\nu}}{\mu\rightarrow e\bar{\nu}\nu}\right]$ & $1.0010 \pm 0.0014$ \\
		$R\left[\frac{\tau\rightarrow \pi\nu}{\pi\rightarrow \mu\bar{\nu}}\right]$& $0.9958 \pm 0.0026$ \\
		$R\left[\frac{\tau\rightarrow K\nu}{K\rightarrow \mu\bar{\nu}}\right]$& $0.9879 \pm 0.0063$ \\
		$R\left[\frac{\tau\rightarrow \mu\nu\bar{\nu}}{\mu\rightarrow e\nu\bar{\nu}}\right]$ & $1.0029 \pm 0.0014$
	\end{tabular}
	\caption{Measurements ratios testing LFU violation in $\tau$ that appear in 
\cref{eq:tauenunu,eq:taumununu2,eq:tauKnu2} from~\cite{HFLAV:2019otj}. The correlations 
 are given in the same reference.
    }
	\label{tab:LFUobs}
\end{table}

\section{Cabibbo angle anomaly}
\label{sec:CAA}

The Cabibbo angle anomaly (CAA) (see Ref.~\cite{Crivellin:2022ctt} for a review) is a deficit in the first-row CKM unitarity~\cite{ParticleDataGroup:2020ssz} 
\begin{align}
	\big|V_{ud}\big|^2+\big|V_{us}\big|^2+\big|V_{ub}\big|^2
	= 0.9985(5).
	\label{eq:CKM}
\end{align} 
Here, $V_{us}$ is most precisely determined from kaon, pion and $\tau$ decays and $V_{ud}$ from beta decays, in particular super allowed and mirrored beta decays. The impact of $V_{ub}$ and its uncertainty its completely negligible.
Explaining the CAA requires $\delta V_{ud}= -0.0075 \pm 0.0025$. The simplified model of \cref{eq:SimplLag} give the NP contributions at the tree level given in~\cref{eq:deltaV}.
Assuming $g^q \leq g^{\ell}_{22}$, which is fixed by the quark charge assignment, the only possibility to achieve a sizable contribution with the right sign is $g^H < g^{\ell}_{11}$. Putting the Higgs on the first site and $\ell_1$ on the second site would achieve this, but lead to a suppressed top Yukawa. Notice that more general realizations with two Higgs doublets, one in the first site and one in the second site, would allow us to find a compromise between the large top Yukawa and an explanation of the CAA. Indeed, the SM Higgs --the Higgs that gets a VEV $v$-- could be a combination of the two Higgses, with $\theta_H$ the mixing angle. Then, $g^H = \cos^2 \theta_H g_2 + \sin^2 \theta_H g_1$,  so an intermediate $g_1 \leq g^H \leq g_2$ is possible. At the same time, the VEV of the second-site Higgs would be $\langle H_2 \rangle = v \cos \theta_H $, which could allow for an $O(1)$ top Yukawa. However, the first-site Higgs would also develop an EW VEV for sizable mixing $\theta_H$, $\langle H_1 \rangle = v \sin \theta_H$, that would spoil the explanation of the flavour hierarchies in the quark sector that our models have.
As an example, we take model 2, and choose the triplet mass at the LHC searches limit $M_{W^{\prime}}\sim 10\,$TeV (see \cref{fig:U2bounds}, and notice that LHC limits do not depend on $g^H$), and a large $g^{\prime}_{2} \sim 4$ to maximize the effect. Then, explaining the deficit of the CAA at one sigma would imply $\sin \theta_H\sim 0.4-0.5$, far from the suppression of $V_{cb}$ and second family Yukawas.

\section{$B_{d,s}-\bar B_{d,s}$ mixing}
\label{sec:BMixing}

Our models generate the operators~\cref{eq:DeltaF2L}, that at the matching scale receive the contribution given in~\cref{eq:DeltaF2Op}.
They contribute to $B_{d,s}$ mixing like
\begin{equation}
\frac{\Delta m_{B_{d,s}}}{\Delta m_{B_{d,s}}^{\rm SM}}=\left| 1+
\frac{C^{\rm NP}_{B_{d,s}}}{C^{\rm SM}_{B_{d,s}}}\right|,
\end{equation}
where, at the EW scale,
\begin{align}
 C_{B_{d}}^{\rm SM}(\mu_{\rm EW})=&\frac{g_L^2}{32 \pi^2 v^2 }(V_{tb}V^*_{td})^2 S_0,\nonumber\\
C_{B_{s}}^{\rm SM}(\mu_{\rm EW})=&\frac{g_L^2}{32 \pi^2 v^2 }(V_{tb}V^*_{ts})^2 S_0,
\end{align}
and $S_0\approx 2.49$~\cite{Buras:1998raa,Buchalla:1995vs}. We have that~\cite{UTfit:2007eik,FlavConstraints}
\begin{equation}
\frac{\Delta m^{\rm Exp}_{B_d}}{\Delta m_{B_d}^{\rm SM}}=1.09\pm 0.09,~~~\frac{\Delta m^{\rm Exp}_{B_s}}{\Delta m_{B_s}^{\rm SM}}=1.10\pm 0.06.
\end{equation}
Including the running to the matching scale $\mu_{\rm UV} \approx 10\,$TeV,
$C^{\rm NP}_{B_s}(\mu_{\rm EW})\approx 0.82\, C^{\rm NP}_{B_s}(\mu_{\rm UV})$~\cite{Fuentes-Martin:2020zaz}, and assuming that the only phases come from the CKM elements, we get the limits reported in \cref{eq:BsM,eq:BdM}.

\section{A protection for $B_s\to \ell^+ \ell^-$}
\label{sec:U1protection}
Let us assume a generic gauge extension of the SM giving a heavy $Z^{\prime}$ associated to some $U(1)^{\prime}$. Working in components of the $SU(2)_L$ doublets, we can write the couplings
\begin{align}
\mathcal{L}\supset  -Z^{\prime}_{\mu} \big( &
g^{H_d} H_d^{*} D^{\mu} H_d+g^q_{23} \bar s_L \gamma^{\mu} b_L +{\rm h.c.} \nonumber\\
&+ 
g^{e_L}_{ii} \bar e^i_L \gamma^{\mu} e^i_L + 
g^{e_R}_{ii} \bar e^i_R \gamma^{\mu} e^i_R\big),
\end{align}
where $H_d$ is the down component of the Higgs getting the VEV, $H=(H_u,H_d)^T$, and the couplings $g$ are proportional to the $U(1)^{\prime}$-charge of each field.
Note that the Higgs coupling will generate a $Z-Z^{\prime}$ mixing when the Higgs gets a VEV $H_d = v/\sqrt{2}$.
When the $Z^{\prime}$ is integrated out, these couplings contribute to the Wilson coefficients of the $bs\ell \ell$ operators defined in ~\cref{eq:O9,eq:O10},
\begin{equation}
\mathcal{L}\supset \tilde C_{9}^{\ell_i} O_{9}^{\ell_i}+\tilde C_{10}^{\ell_i} O_{10}^{\ell_i},
\end{equation}
like
\begin{align}
\tilde C_{9}^{\ell_i\,{\rm NP}}=&- \frac{g^q_{23}}{2M_{Z^{\prime}}^2} [g_{ii}^{e_L}+g_{ii}^{e_R}-(1-4 s_W^2)g^{H_d}],\\
\tilde C_{10}^{\ell_i\,{\rm NP}}=&\, \frac{g^q_{23}}{2M_{Z^{\prime}}^2} (g_{ii}^{e_L}-g^{H_d}-g_{ii}^{e_R}).
\end{align}
We see that $C_{10}^{\ell_i\,{\rm NP}}$, and therefore the contribution to $\mathcal{B}(B_s\to \ell_i^+\ell_i^-)$ (see~\cref{eq:Bsmumu}), generically vanishes when the charges are such that the Yukawa $\bar e^i_L H_d e_R^i$ can be written respecting $U(1)^{\prime}$. This mechanism was also observed in the model presented in Ref.~\cite{Davighi:2023evx}.

\section{LFV processes}
\label{sec:AppLFV}

The LEFT operators
\begin{align}
\mathcal{L}\supset  \sum_{\substack{f=e,u,d\\X,Y=L,R}}
C_{ef}^{V,XY} (\bar e_X\gamma_{\mu} e_X)(\bar f_Y\gamma^{\mu} f_Y) ,
\end{align}
with non-diagonal flavour indices on the $e$ field generate LFV processes. In particular Wilson coefficients $C_{ee}^{V,XY}$ generate LFV three-body decays with branching fractions~\cite{Crivellin:2013hpa}:
\begin{align}
{\cal B}(\ell_{k}\to \ell_j \bar\ell_i \ell_i) &= \frac{M_{\ell_k}^5}{1536 \pi^3 \Gamma_{\ell_k}(1+\delta_{ij})}\nonumber\\
\times\bigg(&\left|[C_{ee}^{V,LL}]_{iijk}\right|^2+\left|[C_{ee}^{V,LR}]_{jkii}\right|^2\nonumber\\
+&\left|[C_{ee}^{V,LR}]_{iijk}\right|^2+\left|[C_{ee}^{V,RR}]_{iijk}\right|^2\bigg),
\end{align}
where $k>j\geq i$.
The experimental bounds on these decays at $90\%$ C.L. are~\cite{HFLAV:2019otj,SINDRUM:1987nra}:
\begin{align}
\mathcal{B}(\tau^-\to \mu^-\mu^+\mu^-)<&\,1.1\times 10^{-8},\label{eq:tau3mu}\\
\mathcal{B}(\tau^-\to \mu^-e^+e^-)<&\,1.1\times 10^{-8},\label{eq:taumu2e}\\
\mathcal{B}(\tau^-\to e^-\mu^+\mu^-)<&\,1.6\times 10^{-8},\label{eq:tau2mue}\\
\mathcal{B}(\tau^-\to e^-e^+e^-)<&\,1.4\times 10^{-8},\label{eq:tau3e}\\
\mathcal{B}(\mu^-\to e^-e^+e^-)<&\,1.0\times 10^{-12}.\label{eq:mu3e}
\end{align}
The Wilson coefficients $C_{eu}^{V,XY},\,C_{ed}^{V,XY}$ generate $\mu \to e$ conversion processes~\cite{Kuno:1999jp}. To compute their contribution, we use the formulas given in Ref.~\cite{Ardu:2024bua}. The current most constraining bound comes from SINDRUM~II experiment with gold nuclei~\cite{SINDRUMII:2006dvw}:
\begin{equation}
{\cal B}(\mu \,{\rm Au}\to e\,{\rm Au}) < 7 \times 10^{-13},
\label{eq:muAu2eAu}
\end{equation}
at $90\%$ C.L.

When a vector triplet with interaction terms of \cref{eq:SimplLag} and LFV couplings,
\begin{equation}
-\mathcal{L}\supset \frac{1}{2}W^{\prime (0)a}_{\mu} \sum_{i\neq j}g^{\ell}_{ij}\bar \ell_{i} \gamma^{\mu} \sigma_a \ell_j ,
\end{equation}
is integrated out, 
we get LFV contributions to $C_{\ell \ell}$, $C_{\ell q}^{(3)}$ and $C_{H \ell}^{(3)}$ Wilson coefficients in SMEFT, that generate 
contributions to the LEFT Wilson coefficients
\begin{align}
[C_{ee}^{V,LL}]_{iijk}=&- \frac{g^{\ell}_{jk}(g^{\ell}_{ii}+g^H(2s_W^2-1))}{4 M_{W^{\prime}}^2},\label{eq:eLeL}\\
[C_{ee}^{V,LR}]_{jkii}=&- \frac{g^{\ell}_{jk}g^H s_W^2}{2 M_{W^{\prime}}^2},\label{eq:eLeR}\\
[C_{eu}^{V,LL}]_{jkii}=&\frac{g^{\ell}_{jk}(g^{q}-g^H(1-4s_W^2/3))}{4 M_{W^{\prime}}^2}\label{eq:eLuL},\\
[C_{ed}^{V,LL}]_{jkii}=&-\frac{g^{\ell}_{jk}(g^{q}+g^H(2s_W^2/3-1))}{4 M_{W^{\prime}}^2},\label{eq:eLdL}\\
[C_{eu}^{V,LR}]_{jkii}=&-2 [C_{ed}^{V,LR}]_{jkii}= \frac{g^{\ell}_{jk}g^H s_W^2}{3 M_{W^{\prime}}^2},\label{eq:eLqR}
\end{align}
where $k>j\geq i$
and we have neglected the running, which at one loop is only due to QED.

\bibliographystyle{JHEP}
\bibliography{references}

\providecommand{\href}[2]{#2}\begingroup\raggedright\begin{thebibliography}{100}

\bibitem{Froggatt:1978nt}
C.~D. Froggatt and H.~B. Nielsen, {\it {Hierarchy of Quark Masses, Cabibbo
  Angles and CP Violation}},  {\em Nucl. Phys. B} {\bf 147} (1979) 277--298.

\bibitem{King:2003rf}
S.~F. King and G.~G. Ross, {\it {Fermion masses and mixing angles from SU (3)
  family symmetry and unification}},  {\em Phys. Lett. B} {\bf 574} (2003)
  239--252, [\href{http://arxiv.org/abs/hep-ph/0307190}{{\tt hep-ph/0307190}}].

\bibitem{Buras:2011wi}
A.~J. Buras, M.~V. Carlucci, L.~Merlo, and E.~Stamou, {\it {Phenomenology of a
  Gauged $SU(3)^3$ Flavour Model}},  {\em JHEP} {\bf 03} (2012) 088,
  [\href{http://arxiv.org/abs/1112.4477}{{\tt arXiv:1112.4477}}].

\bibitem{Kaplan:1991dc}
D.~B. Kaplan, {\it {Flavor at SSC energies: A New mechanism for dynamically
  generated fermion masses}},  {\em Nucl. Phys. B} {\bf 365} (1991) 259--278.

\bibitem{Grossman:1999ra}
Y.~Grossman and M.~Neubert, {\it {Neutrino masses and mixings in
  nonfactorizable geometry}},  {\em Phys. Lett. B} {\bf 474} (2000) 361--371,
  [\href{http://arxiv.org/abs/hep-ph/9912408}{{\tt hep-ph/9912408}}].

\bibitem{Gherghetta:2000qt}
T.~Gherghetta and A.~Pomarol, {\it {Bulk fields and supersymmetry in a slice of
  AdS}},  {\em Nucl. Phys. B} {\bf 586} (2000) 141--162,
  [\href{http://arxiv.org/abs/hep-ph/0003129}{{\tt hep-ph/0003129}}].

\bibitem{Keren-Zur:2012buf}
B.~Keren-Zur, P.~Lodone, M.~Nardecchia, D.~Pappadopulo, R.~Rattazzi, and
  L.~Vecchi, {\it {On Partial Compositeness and the CP asymmetry in charm
  decays}},  {\em Nucl. Phys. B} {\bf 867} (2013) 394--428,
  [\href{http://arxiv.org/abs/1205.5803}{{\tt arXiv:1205.5803}}].

\bibitem{Panico:2016ull}
G.~Panico and A.~Pomarol, {\it {Flavor hierarchies from dynamical scales}},
  {\em JHEP} {\bf 07} (2016) 097, [\href{http://arxiv.org/abs/1603.06609}{{\tt
  arXiv:1603.06609}}].

\bibitem{Berezhiani:1983de}
Z.~G. Berezhiani, {\it {The Weak Mixing Angles in Gauge Models with Horizontal
  Symmetry: A New Approach to Quark and Lepton Masses}},  {\em Phys. Lett. B}
  {\bf 129} (1983) 99--102.

\bibitem{Berezhiani1:1993fg}
Z.~G. Berezhiani and R.~Rattazzi, {\it {Inverse hierarchy approach to fermion
  masses}},  {\em Nucl.Phys. B} {\bf 407} (1993) 249--270,
  [\href{http://arxiv.org/abs/hep-ph/9212245}{{\tt hep-ph/9212245}}].

\bibitem{Barbieri:1994cx}
R.~Barbieri, G.~R. Dvali, and A.~Strumia, {\it {Fermion masses and mixings in a
  flavor symmetric GUT}},  {\em Nucl. Phys. B} {\bf 435} (1995) 102--114,
  [\href{http://arxiv.org/abs/hep-ph/9407239}{{\tt hep-ph/9407239}}].

\bibitem{Dvali:2000ha}
G.~R. Dvali and M.~A. Shifman, {\it {Families as neighbors in extra
  dimension}},  {\em Phys. Lett. B} {\bf 475} (2000) 295--302,
  [\href{http://arxiv.org/abs/hep-ph/0001072}{{\tt hep-ph/0001072}}].

\bibitem{Barbieri:1995uv}
R.~Barbieri, G.~R. Dvali, and L.~J. Hall, {\it {Predictions from a $U(2)$
  flavor symmetry in supersymmetric theories}},  {\em Phys. Lett. B} {\bf 377}
  (1996) 76--82, [\href{http://arxiv.org/abs/hep-ph/9512388}{{\tt
  hep-ph/9512388}}].

\bibitem{Barbieri:2011ci}
R.~Barbieri, G.~Isidori, J.~Jones-Perez, P.~Lodone, and D.~M. Straub, {\it
  {$U(2)$ and Minimal Flavour Violation in Supersymmetry}},  {\em Eur. Phys. J.
  C} {\bf 71} (2011) 1725, [\href{http://arxiv.org/abs/1105.2296}{{\tt
  arXiv:1105.2296}}].

\bibitem{Crivellin:2011fb}
A.~Crivellin, L.~Hofer, and U.~Nierste, {\it {The MSSM with a Softly Broken
  $U(2)^3$ Flavor Symmetry}},  {\em PoS} {\bf EPS-HEP2011} (2011) 145,
  [\href{http://arxiv.org/abs/1111.0246}{{\tt arXiv:1111.0246}}].

\bibitem{Barbieri:2012uh}
R.~Barbieri, D.~Buttazzo, F.~Sala, and D.~M. Straub, {\it {Flavour physics from
  an approximate $U(2)^3$ symmetry}},  {\em JHEP} {\bf 07} (2012) 181,
  [\href{http://arxiv.org/abs/1203.4218}{{\tt arXiv:1203.4218}}].

\bibitem{Buras:2012sd}
A.~J. Buras and J.~Girrbach, {\it {On the Correlations between Flavour
  Observables in Minimal $U(2)^3$ Models}},  {\em JHEP} {\bf 01} (2013) 007,
  [\href{http://arxiv.org/abs/1206.3878}{{\tt arXiv:1206.3878}}].

\bibitem{Calibbi:2019lvs}
L.~Calibbi, A.~Crivellin, F.~Kirk, C.~A. Manzari, and L.~Vernazza, {\it
  {$Z^\prime$ models with less-minimal flavour violation}},  {\em Phys. Rev. D}
  {\bf 101} (2020), no.~9 095003, [\href{http://arxiv.org/abs/1910.00014}{{\tt
  arXiv:1910.00014}}].

\bibitem{Fuentes-Martin:2020bnh}
J.~Fuentes-Mart\'\i{}n and P.~Stangl, {\it {Third-family quark-lepton
  unification with a fundamental composite Higgs}},  {\em Phys. Lett. B} {\bf
  811} (2020) 135953, [\href{http://arxiv.org/abs/2004.11376}{{\tt
  arXiv:2004.11376}}].

\bibitem{Fuentes-Martin:2022xnb}
J.~Fuentes-Martin, G.~Isidori, J.~M. Lizana, N.~Selimovic, and B.~A. Stefanek,
  {\it {Flavor hierarchies, flavor anomalies, and Higgs mass from a warped
  extra dimension}},  {\em Phys. Lett. B} {\bf 834} (2022) 137382,
  [\href{http://arxiv.org/abs/2203.01952}{{\tt arXiv:2203.01952}}].

\bibitem{Georgi:1985hf}
H.~Georgi, {\it {A Tool Kit for Builders of Composite Models}},  {\em Nucl.
  Phys. B} {\bf 266} (1986) 274--284.

\bibitem{Douglas:1996sw}
M.~R. Douglas and G.~W. Moore, {\it {D-branes, quivers, and ALE instantons}},
  \href{http://arxiv.org/abs/hep-th/9603167}{{\tt hep-th/9603167}}.

\bibitem{Arkani-Hamed:2001nha}
N.~Arkani-Hamed, A.~G. Cohen, and H.~Georgi, {\it {(De)constructing
  dimensions}},  {\em Phys.Rev. Lett.} {\bf 86} (2001) 4757--4761,
  [\href{http://arxiv.org/abs/hep-th/0104005}{{\tt hep-th/0104005}}].

\bibitem{Hill:2001ps}
C.~T. Hill, S.~Pokorski, and J.~Wang, {\it {Gauge Invariant Effective
  Lagrangian for Kaluza-Klein Modes}},  {\em Phys. Rev. D} {\bf 64} (2001)
  105005, [\href{http://arxiv.org/abs/hep-th/0104035}{{\tt hep-th/0104035}}].

\bibitem{Cheng:2001vd}
H.-C. Cheng, C.~T. Hill, S.~Pokorski, and J.~Wang, {\it {The Standard Model in
  the Latticized Bulk}},  {\em Phys. Rev. D} {\bf 64} (2001) 065007,
  [\href{http://arxiv.org/abs/hep-th/0104179}{{\tt hep-th/0104179}}].

\bibitem{Chivukula:2013kw}
R.~S. Chivukula, E.~H. Simmons, and N.~Vignaroli, {\it {A Flavorful Top-Coloron
  Model}},  {\em Phys. Rev. D} {\bf 87} (2013), no.~7 075002,
  [\href{http://arxiv.org/abs/1302.1069}{{\tt arXiv:1302.1069}}].

\bibitem{Crivellin:2015lwa}
A.~Crivellin, G.~D'Ambrosio, and J.~Heeck, {\it {Addressing the LHC flavor
  anomalies with horizontal gauge symmetries}},  {\em Phys. Rev. D} {\bf 91}
  (2015), no.~7 075006, [\href{http://arxiv.org/abs/1503.03477}{{\tt
  arXiv:1503.03477}}].

\bibitem{Greljo:2018tuh}
A.~Greljo and B.~A. Stefanek, {\it {Third family quark\textendash{}lepton
  unification at the TeV scale}},  {\em Phys. Lett. B} {\bf 782} (2018)
  131--138, [\href{http://arxiv.org/abs/1802.04274}{{\tt arXiv:1802.04274}}].

\bibitem{Crosas:2022quq}
O.~L. Crosas, G.~Isidori, J.~M. Lizana, N.~Selimovic, and B.~A. Stefanek, {\it
  {Flavor non-universal vector leptoquark imprints in $K\to\pi\nu\bar\nu$ and
  $\Delta F=2$ transitions}},  {\em Phys. Lett. B} {\bf 835} (2022) 137525,
  [\href{http://arxiv.org/abs/2207.00018}{{\tt arXiv:2207.00018}}].

\bibitem{Allwicher:2023aql}
L.~Allwicher, G.~Isidori, J.~M. Lizana, N.~Selimovic, and B.~A. Stefanek, {\it
  {Third-family quark-lepton Unification and electroweak precision tests}},
  {\em JHEP} {\bf 05} (2023) 179, [\href{http://arxiv.org/abs/2302.11584}{{\tt
  arXiv:2302.11584}}].

\bibitem{Bordone:2017bld}
M.~Bordone, C.~Cornella, J.~Fuentes-Martin, and G.~Isidori, {\it {A three-site
  gauge model for flavor hierarchies and flavor anomalies}},  {\em Phys. Lett.
  B} {\bf 779} (2018) 317--323, [\href{http://arxiv.org/abs/1712.01368}{{\tt
  arXiv:1712.01368}}].

\bibitem{Blanke:2018sro}
M.~Blanke and A.~Crivellin, {\it {$B$ Meson Anomalies in a Pati-Salam Model
  within the Randall-Sundrum Background}},  {\em Phys. Rev. Lett.} {\bf 121}
  (2018), no.~1 011801, [\href{http://arxiv.org/abs/1801.07256}{{\tt
  arXiv:1801.07256}}].

\bibitem{Davighi:2023iks}
J.~Davighi and G.~Isidori, {\it {Non-universal gauge interactions addressing
  the inescapable link between Higgs and flavour}},  {\em JHEP} {\bf 07} (2023)
  147, [\href{http://arxiv.org/abs/2303.01520}{{\tt arXiv:2303.01520}}].

\bibitem{FernandezNavarro:2023rhv}
M.~Fern\'andez~Navarro and S.~F. King, {\it {Tri-hypercharge: a separate gauged
  weak hypercharge for each fermion family as the origin of flavour}},  {\em
  JHEP} {\bf 08} (2023) 020, [\href{http://arxiv.org/abs/2305.07690}{{\tt
  arXiv:2305.07690}}].

\bibitem{Davighi:2023evx}
J.~Davighi and B.~A. Stefanek, {\it {Deconstructed Hypercharge: A Natural Model
  of Flavour}},  \href{http://arxiv.org/abs/2305.16280}{{\tt
  arXiv:2305.16280}}.

\bibitem{FernandezNavarro:2023hrf}
M.~Fern\'andez~Navarro, S.~F. King, and A.~Vicente, {\it {Tri-unification: a
  separate $SU(5)$ for each fermion family}},
  \href{http://arxiv.org/abs/2311.05683}{{\tt arXiv:2311.05683}}.

\bibitem{Barbieri:2023qpf}
R.~Barbieri and G.~Isidori, {\it {Minimal flavour deconstruction}},
  \href{http://arxiv.org/abs/2312.14004}{{\tt arXiv:2312.14004}}.

\bibitem{Greljo:2023bix}
A.~Greljo and A.~E. Thomsen, {\it {Rising Through the Ranks: Flavor Hierarchies
  from a Gauged $SU(2)$ Symmetry}},
  \href{http://arxiv.org/abs/2309.11547}{{\tt arXiv:2309.11547}}.

\bibitem{Li:1981nk}
X.~Li and E.~Ma, {\it {Gauge Model of Generation Nonuniversality}},  {\em Phys.
  Rev. Lett.} {\bf 47} (1981) 1788.

\bibitem{Muller:1996dj}
D.~J. Muller and S.~Nandi, {\it {Top flavor: A Separate $SU(2)$ for the third
  family}},  {\em Phys. Lett. B} {\bf 383} (1996) 345--350,
  [\href{http://arxiv.org/abs/hep-ph/9602390}{{\tt hep-ph/9602390}}].

\bibitem{Malkawi:1996fs}
E.~Malkawi, T.~M.~P. Tait, and C.~P. Yuan, {\it {A Model of strong flavor
  dynamics for the top quark}},  {\em Phys. Lett. B} {\bf 385} (1996) 304--310,
  [\href{http://arxiv.org/abs/hep-ph/9603349}{{\tt hep-ph/9603349}}].

\bibitem{Shu:2006mm}
J.~Shu, T.~M.~P. Tait, and C.~E.~M. Wagner, {\it {Baryogenesis from an Earlier
  Phase Transition}},  {\em Phys. Rev. D} {\bf 75} (2007) 063510,
  [\href{http://arxiv.org/abs/hep-ph/0610375}{{\tt hep-ph/0610375}}].

\bibitem{Chiang:2009kb}
C.-W. Chiang, N.~G. Deshpande, X.-G. He, and J.~Jiang, {\it {The Family
  $SU(2)_l\times SU(2)_h\times U(1)$ Model}},  {\em Phys. Rev. D} {\bf 81}
  (2010) 015006, [\href{http://arxiv.org/abs/0911.1480}{{\tt
  arXiv:0911.1480}}].

\bibitem{Hsieh:2010zr}
K.~Hsieh, K.~Schmitz, J.-H. Yu, and C.~P. Yuan, {\it {Global Analysis of
  General $SU(2) \times SU(2) \times U(1)$ Models with Precision Data}},  {\em
  Phys. Rev. D} {\bf 82} (2010) 035011,
  [\href{http://arxiv.org/abs/1003.3482}{{\tt arXiv:1003.3482}}].

\bibitem{ATLAS:2024tzc}
{\bf ATLAS} Collaboration, G.~Aad et~al., {\it {Search for high-mass resonances
  in final states with a $\tau$-lepton and missing transverse momentum with the
  ATLAS detector}},  \href{http://arxiv.org/abs/2402.16576}{{\tt
  arXiv:2402.16576}}.

\bibitem{Antusch:2023shi}
S.~Antusch, A.~Greljo, B.~A. Stefanek, and A.~E. Thomsen, {\it {$U(2)$ is Right
  for Leptons and Left for Quarks}},
  \href{http://arxiv.org/abs/2311.09288}{{\tt arXiv:2311.09288}}.

\bibitem{deBlas:2012qp}
J.~de~Blas, J.~M. Lizana, and M.~Perez-Victoria, {\it {Combining searches of
  $Z'$ and $W'$ bosons}},  {\em JHEP} {\bf 01} (2013) 166,
  [\href{http://arxiv.org/abs/1211.2229}{{\tt arXiv:1211.2229}}].

\bibitem{Pappadopulo:2014qza}
D.~Pappadopulo, A.~Thamm, R.~Torre, and A.~Wulzer, {\it {Heavy Vector Triplets:
  Bridging Theory and Data}},  {\em JHEP} {\bf 09} (2014) 060,
  [\href{http://arxiv.org/abs/1402.4431}{{\tt arXiv:1402.4431}}].

\bibitem{Davighi:2023xqn}
J.~Davighi, A.~Gosnay, D.~J. Miller, and S.~Renner, {\it {Phenomenology of a
  Deconstructed Electroweak Force}},
  \href{http://arxiv.org/abs/2312.13346}{{\tt arXiv:2312.13346}}.

\bibitem{Capdevila:2023yhq}
B.~Capdevila, A.~Crivellin, and J.~Matias, {\it {Review of Semileptonic $B$
  Anomalies}},  {\em Eur. Phys. J. ST} {\bf 1} (2023) 20,
  [\href{http://arxiv.org/abs/2309.01311}{{\tt arXiv:2309.01311}}].

\bibitem{Branco:2011iw}
G.~C. Branco, P.~M. Ferreira, L.~Lavoura, M.~N. Rebelo, M.~Sher, and J.~P.
  Silva, {\it {Theory and phenomenology of two-Higgs-doublet models}},  {\em
  Phys. Rept.} {\bf 516} (2012) 1--102,
  [\href{http://arxiv.org/abs/1106.0034}{{\tt arXiv:1106.0034}}].

\bibitem{Fuentes-Martin:2020pww}
J.~Fuentes-Martin, G.~Isidori, J.~Pag\`es, and B.~A. Stefanek, {\it {Flavor
  non-universal Pati-Salam unification and neutrino masses}},  {\em Phys. Lett.
  B} {\bf 820} (2021) 136484, [\href{http://arxiv.org/abs/2012.10492}{{\tt
  arXiv:2012.10492}}].

\bibitem{Fuentes-Martin:2024fpx}
J.~Fuentes-Mart\'\i{}n and J.~M. Lizana, {\it {Deconstructing flavor
  anomalously}},  {\em JHEP} {\bf 07} (2024) 117,
  [\href{http://arxiv.org/abs/2402.09507}{{\tt arXiv:2402.09507}}].

\bibitem{Clark:2016jgm}
D.~B. Clark, E.~Godat, and F.~I. Olness, {\it {ManeParse : A Mathematica reader
  for Parton Distribution Functions}},  {\em Comput. Phys. Commun.} {\bf 216}
  (2017) 126--137, [\href{http://arxiv.org/abs/1605.08012}{{\tt
  arXiv:1605.08012}}].

\bibitem{ATLAS:2019lsy}
{\bf ATLAS} Collaboration, G.~Aad et~al., {\it {Search for a heavy charged
  boson in events with a charged lepton and missing transverse momentum from
  $pp$ collisions at $\sqrt{s} = 13$ TeV with the ATLAS detector}},  {\em Phys.
  Rev. D} {\bf 100} (2019), no.~5 052013,
  [\href{http://arxiv.org/abs/1906.05609}{{\tt arXiv:1906.05609}}].

\bibitem{CMS:2021ctt}
{\bf CMS} Collaboration, A.~M. Sirunyan et~al., {\it {Search for resonant and
  nonresonant new phenomena in high-mass dilepton final states at $ \sqrt{s} $
  = 13 TeV}},  {\em JHEP} {\bf 07} (2021) 208,
  [\href{http://arxiv.org/abs/2103.02708}{{\tt arXiv:2103.02708}}].

\bibitem{ATLAS:2021bjk}
{\bf ATLAS} Collaboration, {\it {Search for high-mass resonances in final
  states with a tau lepton and missing transverse momentum with the ATLAS
  detector}}, .

\bibitem{ATLAS:2019fgd}
{\bf ATLAS} Collaboration, G.~Aad et~al., {\it {Search for new resonances in
  mass distributions of jet pairs using 139 fb$^{-1}$ of $pp$ collisions at
  $\sqrt{s}=13$ TeV with the ATLAS detector}},  {\em JHEP} {\bf 03} (2020) 145,
  [\href{http://arxiv.org/abs/1910.08447}{{\tt arXiv:1910.08447}}].

\bibitem{ATLAS:2017eiz}
{\bf ATLAS} Collaboration, M.~Aaboud et~al., {\it {Search for additional heavy
  neutral Higgs and gauge bosons in the ditau final state produced in 36
  fb$^{-1}$ of pp collisions at $ \sqrt{s}=13 $ TeV with the ATLAS detector}},
  {\em JHEP} {\bf 01} (2018) 055, [\href{http://arxiv.org/abs/1709.07242}{{\tt
  arXiv:1709.07242}}].

\bibitem{Allwicher:2022mcg}
L.~Allwicher, D.~A. Faroughy, F.~Jaffredo, O.~Sumensari, and F.~Wilsch, {\it
  {HighPT: A tool for~ high-$p_T$ Drell-Yan tails beyond the standard model}},
  {\em Comput. Phys. Commun.} {\bf 289} (2023) 108749,
  [\href{http://arxiv.org/abs/2207.10756}{{\tt arXiv:2207.10756}}].

\bibitem{ATLAS:2019itm}
{\bf ATLAS} Collaboration, M.~Aaboud et~al., {\it {Search for low-mass
  resonances decaying into two jets and produced in association with a photon
  using $pp$ collisions at $\sqrt{s} = 13$ TeV with the ATLAS detector}},  {\em
  Phys. Lett. B} {\bf 795} (2019) 56--75,
  [\href{http://arxiv.org/abs/1901.10917}{{\tt arXiv:1901.10917}}].

\bibitem{Bartocci:2023nvp}
R.~Bartocci, A.~Biek\"otter, and T.~Hurth, {\it {A global analysis of the SMEFT
  under the minimal MFV assumption}},
  \href{http://arxiv.org/abs/2311.04963}{{\tt arXiv:2311.04963}}.

\bibitem{Breso-Pla:2021qoe}
V.~Bres\'o-Pla, A.~Falkowski, and M.~Gonz\'alez-Alonso, {\it {$A_{FB}$ in the
  SMEFT: precision $Z$ physics at the LHC}},  {\em JHEP} {\bf 08} (2021) 021,
  [\href{http://arxiv.org/abs/2103.12074}{{\tt arXiv:2103.12074}}].

\bibitem{ParticleDataGroup:2018ovx}
{\bf Particle Data Group} Collaboration, M.~Tanabashi et~al., {\it {Review of
  Particle Physics}},  {\em Phys. Rev. D} {\bf 98} (2018), no.~3 030001.

\bibitem{CDF:2022hxs}
{\bf CDF} Collaboration, T.~Aaltonen et~al., {\it {High-precision measurement
  of the $W$ boson mass with the CDF II detector}},  {\em Science} {\bf 376}
  (2022), no.~6589 170--176.

\bibitem{ATLAS:2023fsi}
{\bf ATLAS} Collaboration, {\it {Improved $W$ boson Mass Measurement using 7
  TeV Proton-Proton Collisions with the ATLAS Detector}}, .

\bibitem{Bagnaschi:2022whn}
E.~Bagnaschi, J.~Ellis, M.~Madigan, K.~Mimasu, V.~Sanz, and T.~You, {\it {SMEFT
  analysis of $m_{W}$}},  {\em JHEP} {\bf 08} (2022) 308,
  [\href{http://arxiv.org/abs/2204.05260}{{\tt arXiv:2204.05260}}].

\bibitem{Falkowski:2015krw}
A.~Falkowski and K.~Mimouni, {\it {Model independent constraints on four-lepton
  operators}},  {\em JHEP} {\bf 02} (2016) 086,
  [\href{http://arxiv.org/abs/1511.07434}{{\tt arXiv:1511.07434}}].

\bibitem{Allanach:2023uxz}
B.~Allanach and A.~Mullin, {\it {Plan B: new $Z^{\prime}$ models for $b
  \rightarrow{} s\ell^{+}\ell^{-}$ anomalies}},  {\em JHEP} {\bf 09} (2023)
  173, [\href{http://arxiv.org/abs/2306.08669}{{\tt arXiv:2306.08669}}].

\bibitem{ALEPH:2013dgf}
{\bf ALEPH, DELPHI, L3, OPAL, LEP Electroweak} Collaboration, S.~Schael et~al.,
  {\it {Electroweak Measurements in Electron-Positron Collisions at
  W-Boson-Pair Energies at LEP}},  {\em Phys. Rept.} {\bf 532} (2013) 119--244,
  [\href{http://arxiv.org/abs/1302.3415}{{\tt arXiv:1302.3415}}].

\bibitem{Allwicher:2023shc}
L.~Allwicher, C.~Cornella, G.~Isidori, and B.~A. Stefanek, {\it {New physics in
  the third generation. A comprehensive SMEFT analysis and future prospects}},
  {\em JHEP} {\bf 03} (2024) 049, [\href{http://arxiv.org/abs/2311.00020}{{\tt
  arXiv:2311.00020}}].

\bibitem{Crivellin:2020lzu}
A.~Crivellin and M.~Hoferichter, {\it {\ensuremath{\beta} Decays as Sensitive
  Probes of Lepton Flavor Universality}},  {\em Phys. Rev. Lett.} {\bf 125}
  (2020), no.~11 111801, [\href{http://arxiv.org/abs/2002.07184}{{\tt
  arXiv:2002.07184}}].

\bibitem{Capdevila:2020rrl}
B.~Capdevila, A.~Crivellin, C.~A. Manzari, and M.~Montull, {\it {Explaining
  $b\to s\ell^+\ell^-$ and the Cabibbo angle anomaly with a vector triplet}},
  {\em Phys. Rev. D} {\bf 103} (2021), no.~1 015032,
  [\href{http://arxiv.org/abs/2005.13542}{{\tt arXiv:2005.13542}}].

\bibitem{Morrissey:2005uza}
D.~E. Morrissey, T.~M.~P. Tait, and C.~E.~M. Wagner, {\it {Proton lifetime and
  baryon number violating signatures at the CERN LHC in gauge extended
  models}},  {\em Phys. Rev. D} {\bf 72} (2005) 095003,
  [\href{http://arxiv.org/abs/hep-ph/0508123}{{\tt hep-ph/0508123}}].

\bibitem{Fuentes-Martin:2014fxa}
J.~Fuentes-Martin, J.~Portoles, and P.~Ruiz-Femenia, {\it {Instanton-mediated
  baryon number violation in non-universal gauge extended models}},  {\em JHEP}
  {\bf 01} (2015) 134, [\href{http://arxiv.org/abs/1411.2471}{{\tt
  arXiv:1411.2471}}].

\bibitem{FlavConstraints0}
``{Flavor Constraints on new physics}.''
  \url{https://agenda.infn.it/event/14377/contributions/24434/attachments/17481/19830/silvestriniLaThuile.pdf}.
\newblock La Thuile 2018.

\bibitem{UTfit:2007eik}
{\bf UTfit} Collaboration, M.~Bona et~al., {\it {Model-independent constraints
  on $\Delta F=2$ operators and the scale of new physics}},  {\em JHEP} {\bf
  03} (2008) 049, [\href{http://arxiv.org/abs/0707.0636}{{\tt
  arXiv:0707.0636}}].

\bibitem{Bobeth:2013uxa}
C.~Bobeth, M.~Gorbahn, T.~Hermann, M.~Misiak, E.~Stamou, and M.~Steinhauser,
  {\it {$B_{s,d} \to \ell^+ \ell^-$ in the Standard Model with Reduced
  Theoretical Uncertainty}},  {\em Phys. Rev. Lett.} {\bf 112} (2014) 101801,
  [\href{http://arxiv.org/abs/1311.0903}{{\tt arXiv:1311.0903}}].

\bibitem{Beneke:2019slt}
M.~Beneke, C.~Bobeth, and R.~Szafron, {\it {Power-enhanced leading-logarithmic
  QED corrections to $B_q \to \mu^+\mu^-$}},  {\em JHEP} {\bf 10} (2019) 232,
  [\href{http://arxiv.org/abs/1908.07011}{{\tt arXiv:1908.07011}}]. [Erratum:
  JHEP 11, 099 (2022)].

\bibitem{Neshatpour:2022pvg}
S.~Neshatpour, T.~Hurth, F.~Mahmoudi, and D.~Martinez~Santos, {\it {Neutral
  Current $B$-Decay Anomalies}},  {\em Springer Proc. Phys.} {\bf 292} (2023)
  11--21, [\href{http://arxiv.org/abs/2210.07221}{{\tt arXiv:2210.07221}}].

\bibitem{Altmannshofer:2021qrr}
W.~Altmannshofer and P.~Stangl, {\it {New physics in rare $B$ decays after
  Moriond 2021}},  {\em Eur. Phys. J. C} {\bf 81} (2021), no.~10 952,
  [\href{http://arxiv.org/abs/2103.13370}{{\tt arXiv:2103.13370}}].

\bibitem{LHCb:2022vje}
{\bf LHCb} Collaboration, R.~Aaij et~al., {\it {Measurement of lepton
  universality parameters in $B^+\to K^+\ell^+\ell^-$ and $B^0\to
  K^{*0}\ell^+\ell^-$ decays}},  {\em Phys. Rev. D} {\bf 108} (2023), no.~3
  032002, [\href{http://arxiv.org/abs/2212.09153}{{\tt arXiv:2212.09153}}].

\bibitem{LHCb:2022qnv}
{\bf LHCb} Collaboration, R.~Aaij et~al., {\it {Test of lepton universality in
  $b \rightarrow s \ell^+ \ell^-$ decays}},  {\em Phys. Rev. Lett.} {\bf 131}
  (2023), no.~5 051803, [\href{http://arxiv.org/abs/2212.09152}{{\tt
  arXiv:2212.09152}}].

\bibitem{Alguero:2023jeh}
M.~Alguer\'o, A.~Biswas, B.~Capdevila, S.~Descotes-Genon, J.~Matias, and
  M.~Novoa-Brunet, {\it {To (b)e or not to (b)e: no electrons at LHCb}},  {\em
  Eur. Phys. J. C} {\bf 83} (2023), no.~7 648,
  [\href{http://arxiv.org/abs/2304.07330}{{\tt arXiv:2304.07330}}].

\bibitem{Capdevila:2023hiv}
B.~Capdevila, {\it {Status of the global $b\to s\ell^+\ell^-$ fits}},  {\em
  PoS} {\bf FPCP2023} (10, 2023) 010.

\bibitem{Descotes-Genon:2012isb}
S.~Descotes-Genon, J.~Matias, M.~Ramon, and J.~Virto, {\it {Implications from
  clean observables for the binned analysis of $B -> K*\mu^+\mu^-$ at large
  recoil}},  {\em JHEP} {\bf 01} (2013) 048,
  [\href{http://arxiv.org/abs/1207.2753}{{\tt arXiv:1207.2753}}].

\bibitem{Parrott:2022rgu}
{\bf (HPQCD collaboration)\textsection{}, HPQCD} Collaboration, W.~G. Parrott,
  C.~Bouchard, and C.~T.~H. Davies, {\it {B\textrightarrow{}K and
  D\textrightarrow{}K form factors from fully relativistic lattice QCD}},  {\em
  Phys. Rev. D} {\bf 107} (2023), no.~1 014510,
  [\href{http://arxiv.org/abs/2207.12468}{{\tt arXiv:2207.12468}}].

\bibitem{Greljo:2022jac}
A.~Greljo, J.~Salko, A.~Smolkovi\v{c}, and P.~Stangl, {\it {Rare $b$ decays
  meet high-mass Drell-Yan}},  {\em JHEP} {\bf 05} (2023) 087,
  [\href{http://arxiv.org/abs/2212.10497}{{\tt arXiv:2212.10497}}].

\bibitem{Ciuchini:2022wbq}
M.~Ciuchini, M.~Fedele, E.~Franco, A.~Paul, L.~Silvestrini, and M.~Valli, {\it
  {Constraints on lepton universality violation from rare B decays}},  {\em
  Phys. Rev. D} {\bf 107} (2023), no.~5 055036,
  [\href{http://arxiv.org/abs/2212.10516}{{\tt arXiv:2212.10516}}].

\bibitem{Hurth:2023jwr}
T.~Hurth, F.~Mahmoudi, and S.~Neshatpour, {\it {B anomalies in the post-RK(*)
  era}},  {\em Phys. Rev. D} {\bf 108} (2023), no.~11 115037,
  [\href{http://arxiv.org/abs/2310.05585}{{\tt arXiv:2310.05585}}].

\bibitem{Khodjamirian:2010vf}
A.~Khodjamirian, T.~Mannel, A.~A. Pivovarov, and Y.~M. Wang, {\it {Charm-loop
  effect in $B \to K^{(*)} \ell^{+} \ell^{-}$ and $B\to K^*\gamma$}},  {\em
  JHEP} {\bf 09} (2010) 089, [\href{http://arxiv.org/abs/1006.4945}{{\tt
  arXiv:1006.4945}}].

\bibitem{Gubernari:2020eft}
N.~Gubernari, D.~van Dyk, and J.~Virto, {\it {Non-local matrix elements in
  $B_{(s)}\to \{K^{(*)},\phi\}\ell^+\ell^-$}},  {\em JHEP} {\bf 02} (2021) 088,
  [\href{http://arxiv.org/abs/2011.09813}{{\tt arXiv:2011.09813}}].

\bibitem{Buras:1998raa}
A.~J. Buras, {\it {Weak Hamiltonian, CP violation and rare decays}},  in {\em
  {Les Houches Summer School in Theoretical Physics, Session 68: Probing the
  Standard Model of Particle Interactions}}, pp.~281--539, 6, 1998.
\newblock \href{http://arxiv.org/abs/hep-ph/9806471}{{\tt hep-ph/9806471}}.

\bibitem{Buras:2015qea}
A.~J. Buras, D.~Buttazzo, J.~Girrbach-Noe, and R.~Knegjens, {\it {$ {K}^{+}\to
  {\pi}^{+}\nu \overline{\nu} $ and $ {K}_L\to {\pi}^0\nu \overline{\nu} $ in
  the Standard Model: status and perspectives}},  {\em JHEP} {\bf 11} (2015)
  033, [\href{http://arxiv.org/abs/1503.02693}{{\tt arXiv:1503.02693}}].

\bibitem{NA62:2021zjw}
{\bf NA62} Collaboration, E.~Cortina~Gil et~al., {\it {Measurement of the very
  rare K$^{+}$\textrightarrow{}$ {\pi}^{+}\nu \overline{\nu} $ decay}},  {\em
  JHEP} {\bf 06} (2021) 093, [\href{http://arxiv.org/abs/2103.15389}{{\tt
  arXiv:2103.15389}}].

\bibitem{Anzivino:2023bhp}
G.~Anzivino et~al., {\it {Workshop summary -- Kaons@CERN 2023}},  in {\em
  {Kaons@CERN 2023}}, 11, 2023.
\newblock \href{http://arxiv.org/abs/2311.02923}{{\tt arXiv:2311.02923}}.

\bibitem{Ahdida:2023okr}
C.~Ahdida et~al., {\it {Post-LS3 Experimental Options in ECN3}},
  \href{http://arxiv.org/abs/2310.17726}{{\tt arXiv:2310.17726}}.

\bibitem{Aoki:2021cqa}
K.~Aoki et~al., {\it {Extension of the J-PARC Hadron Experimental Facility:
  Third White Paper}},  \href{http://arxiv.org/abs/2110.04462}{{\tt
  arXiv:2110.04462}}.

\bibitem{Belle-II:2018jsg}
{\bf Belle-II} Collaboration, W.~Altmannshofer et~al., {\it {The Belle II
  Physics Book}},  {\em PTEP} {\bf 2019} (2019), no.~12 123C01,
  [\href{http://arxiv.org/abs/1808.10567}{{\tt arXiv:1808.10567}}]. [Erratum:
  PTEP 2020, 029201 (2020)].

\bibitem{BKnunu}
S.~Glazov, ``{Belle II physics highlights}.''
  \url{https://indico.desy.de/event/34916/contributions/149769/attachments/84417/111854/Belle%20II%20highlights.pdf}.
\newblock EPS-HEP2023 conference.

\bibitem{Altmannshofer:2023hkn}
W.~Altmannshofer, A.~Crivellin, H.~Haigh, G.~Inguglia, and J.~Martin~Camalich,
  {\it {Light New Physics in $B\to K^{(*)}\nu\bar\nu$?}},
  \href{http://arxiv.org/abs/2311.14629}{{\tt arXiv:2311.14629}}.

\bibitem{HFLAV:2022esi}
{\bf HFLAV} Collaboration, Y.~S. Amhis et~al., {\it {Averages of $b$-hadron,
  $c$-hadron, and \ensuremath{\tau}-lepton properties as of 2021}},  {\em Phys.
  Rev. D} {\bf 107} (2023), no.~5 052008,
  [\href{http://arxiv.org/abs/2206.07501}{{\tt arXiv:2206.07501}}].

\bibitem{DAmbrosio:2002vsn}
G.~D'Ambrosio, G.~F. Giudice, G.~Isidori, and A.~Strumia, {\it {Minimal flavor
  violation: An Effective field theory approach}},  {\em Nucl. Phys. B} {\bf
  645} (2002) 155--187, [\href{http://arxiv.org/abs/hep-ph/0207036}{{\tt
  hep-ph/0207036}}].

\bibitem{EuropeanStrategyforParticlePhysicsPreparatoryGroup:2019qin}
R.~K. Ellis et~al., {\it {Physics Briefing Book}: {Input for the European
  Strategy for Particle Physics Update 2020}},
  \href{http://arxiv.org/abs/1910.11775}{{\tt arXiv:1910.11775}}.

\bibitem{Lizana:2023kei}
J.~M. Lizana, J.~Matias, and B.~A. Stefanek, {\it {Explaining the $
  {B}_{d,s}\to {K}^{\left(\ast \right)}{\overline{K}}^{\left(\ast \right)} $
  non-leptonic puzzle and charged-current B-anomalies via scalar leptoquarks}},
   {\em JHEP} {\bf 09} (2023) 114, [\href{http://arxiv.org/abs/2306.09178}{{\tt
  arXiv:2306.09178}}].

\bibitem{Hesketh:2022wgw}
{\bf Mu3e} Collaboration, G.~Hesketh, S.~Hughes, A.-K. Perrevoort, and
  N.~Rompotis, {\it {The Mu3e Experiment}},  in {\em {Snowmass 2021}}, 4, 2022.
\newblock \href{http://arxiv.org/abs/2204.00001}{{\tt arXiv:2204.00001}}.

\bibitem{LHCb:2018roe}
{\bf LHCb} Collaboration, R.~Aaij et~al., {\it {Physics case for an LHCb
  Upgrade II - Opportunities in flavour physics, and beyond, in the HL-LHC
  era}},  \href{http://arxiv.org/abs/1808.08865}{{\tt arXiv:1808.08865}}.

\bibitem{COMET:2009qeh}
{\bf COMET} Collaboration, Y.~G. Cui et~al., {\it {Conceptual design report for
  experimental search for lepton flavor violating mu- - e- conversion at
  sensitivity of 10**(-16) with a slow-extracted bunched proton beam (COMET)}},
  .

\bibitem{Mu2e:2014fns}
{\bf Mu2e} Collaboration, L.~Bartoszek et~al., {\it {Mu2e Technical Design
  Report}},  \href{http://arxiv.org/abs/1501.05241}{{\tt arXiv:1501.05241}}.

\bibitem{Cerri:2018ypt}
A.~Cerri et~al., {\it {Report from Working Group 4}: {Opportunities in Flavour
  Physics at the HL-LHC and HE-LHC}},  {\em CERN Yellow Rep. Monogr.} {\bf 7}
  (2019) 867--1158, [\href{http://arxiv.org/abs/1812.07638}{{\tt
  arXiv:1812.07638}}].

\bibitem{FCC:2018evy}
{\bf FCC} Collaboration, A.~Abada et~al., {\it {FCC-ee: The Lepton Collider}:
  {Future Circular Collider Conceptual Design Report Volume 2}},  {\em Eur.
  Phys. J. ST} {\bf 228} (2019), no.~2 261--623.

\bibitem{FCC:2018byv}
{\bf FCC} Collaboration, A.~Abada et~al., {\it {FCC Physics Opportunities}:
  {Future Circular Collider Conceptual Design Report Volume 1}},  {\em Eur.
  Phys. J. C} {\bf 79} (2019), no.~6 474.

\bibitem{CEPCStudyGroup:2018ghi}
{\bf CEPC Study Group} Collaboration, M.~Dong et~al., {\it {CEPC Conceptual
  Design Report: Volume 2 - Physics \& Detector}},
  \href{http://arxiv.org/abs/1811.10545}{{\tt arXiv:1811.10545}}.

\bibitem{deBlas:2022ofj}
J.~de~Blas, Y.~Du, C.~Grojean, J.~Gu, V.~Miralles, M.~E. Peskin, J.~Tian,
  M.~Vos, and E.~Vryonidou, {\it {Global SMEFT Fits at Future Colliders}},  in
  {\em {Snowmass 2021}}, 6, 2022.
\newblock \href{http://arxiv.org/abs/2206.08326}{{\tt arXiv:2206.08326}}.

\bibitem{Dam:2018rfz}
M.~Dam, {\it {Tau-lepton Physics at the FCC-ee circular e$^+$e$^-$ Collider}},
  {\em SciPost Phys. Proc.} {\bf 1} (2019) 041,
  [\href{http://arxiv.org/abs/1811.09408}{{\tt arXiv:1811.09408}}].

\bibitem{Davighi:2022fer}
J.~Davighi and J.~Tooby-Smith, {\it {Electroweak flavour unification}},  {\em
  JHEP} {\bf 09} (2022) 193, [\href{http://arxiv.org/abs/2201.07245}{{\tt
  arXiv:2201.07245}}].

\bibitem{Davighi:2022bqf}
J.~Davighi, G.~Isidori, and M.~Pesut, {\it {Electroweak-flavour and
  quark-lepton unification: a family non-universal path}},  {\em JHEP} {\bf 04}
  (2023) 030, [\href{http://arxiv.org/abs/2212.06163}{{\tt arXiv:2212.06163}}].

\bibitem{Grzadkowski:2010es}
B.~Grzadkowski, M.~Iskrzynski, M.~Misiak, and J.~Rosiek, {\it {Dimension-Six
  Terms in the Standard Model Lagrangian}},  {\em JHEP} {\bf 10} (2010) 085,
  [\href{http://arxiv.org/abs/1008.4884}{{\tt arXiv:1008.4884}}].

\bibitem{Fuentes-Martin:2022jrf}
J.~Fuentes-Mart\'\i{}n, M.~K\"onig, J.~Pag\`es, A.~E. Thomsen, and F.~Wilsch,
  {\it {A proof of concept for matchete: an automated tool for matching
  effective theories}},  {\em Eur. Phys. J. C} {\bf 83} (2023), no.~7 662,
  [\href{http://arxiv.org/abs/2212.04510}{{\tt arXiv:2212.04510}}].

\bibitem{CMS:2022nty}
{\bf CMS} Collaboration, A.~Tumasyan et~al., {\it {Inclusive nonresonant
  multilepton probes of new phenomena at $\sqrt s$=13\,\,TeV}},  {\em Phys.
  Rev. D} {\bf 105} (2022), no.~11 112007,
  [\href{http://arxiv.org/abs/2202.08676}{{\tt arXiv:2202.08676}}].

\bibitem{Crivellin:2020ebi}
A.~Crivellin, F.~Kirk, C.~A. Manzari, and M.~Montull, {\it {Global Electroweak
  Fit and Vector-Like Leptons in Light of the Cabibbo Angle Anomaly}},  {\em
  JHEP} {\bf 12} (2020) 166, [\href{http://arxiv.org/abs/2008.01113}{{\tt
  arXiv:2008.01113}}].

\bibitem{HFLAV:2019otj}
{\bf HFLAV} Collaboration, Y.~S. Amhis et~al., {\it {Averages of $b$-hadron,
  $c$-hadron, and $\tau $-lepton properties as of 2018}},  {\em Eur. Phys. J.
  C} {\bf 81} (2021), no.~3 226, [\href{http://arxiv.org/abs/1909.12524}{{\tt
  arXiv:1909.12524}}].

\bibitem{Crivellin:2022ctt}
A.~Crivellin, {\it {Explaining the Cabibbo Angle Anomaly}},  7, 2022.
\newblock \href{http://arxiv.org/abs/2207.02507}{{\tt arXiv:2207.02507}}.

\bibitem{ParticleDataGroup:2020ssz}
{\bf Particle Data Group} Collaboration, P.~A. Zyla et~al., {\it {Review of
  Particle Physics}},  {\em PTEP} {\bf 2020} (2020), no.~8 083C01.

\bibitem{Buchalla:1995vs}
G.~Buchalla, A.~J. Buras, and M.~E. Lautenbacher, {\it {Weak decays beyond
  leading logarithms}},  {\em Rev. Mod. Phys.} {\bf 68} (1996) 1125--1144,
  [\href{http://arxiv.org/abs/hep-ph/9512380}{{\tt hep-ph/9512380}}].

\bibitem{FlavConstraints}
``{Global fits of the Unitarity Triangle within the Standard Model and beyond.
  Updates from the UTfit collaboration}.''
  \url{https://indico.desy.de/event/34916/contributions/147139/attachments/84166/111441/EPS2023_UTfit.pdf}.
\newblock EPS-HEP2023 conference.

\bibitem{Fuentes-Martin:2020zaz}
J.~Fuentes-Martin, P.~Ruiz-Femenia, A.~Vicente, and J.~Virto, {\it {DsixTools
  2.0: The Effective Field Theory Toolkit}},  {\em Eur. Phys. J. C} {\bf 81}
  (2021), no.~2 167, [\href{http://arxiv.org/abs/2010.16341}{{\tt
  arXiv:2010.16341}}].

\bibitem{Crivellin:2013hpa}
A.~Crivellin, S.~Najjari, and J.~Rosiek, {\it {Lepton Flavor Violation in the
  Standard Model with general Dimension-Six Operators}},  {\em JHEP} {\bf 04}
  (2014) 167, [\href{http://arxiv.org/abs/1312.0634}{{\tt arXiv:1312.0634}}].

\bibitem{SINDRUM:1987nra}
{\bf SINDRUM} Collaboration, U.~Bellgardt et~al., {\it {Search for the Decay
  $\mu^+ \to e^+ e^+ e^-$}},  {\em Nucl. Phys. B} {\bf 299} (1988) 1--6.

\bibitem{Kuno:1999jp}
Y.~Kuno and Y.~Okada, {\it {Muon decay and physics beyond the standard model}},
   {\em Rev. Mod. Phys.} {\bf 73} (2001) 151--202,
  [\href{http://arxiv.org/abs/hep-ph/9909265}{{\tt hep-ph/9909265}}].

\bibitem{Ardu:2024bua}
M.~Ardu, S.~Davidson, and S.~Lavignac, {\it {Constraining new physics models
  from $\mu \rightarrow e $ observables in bottom-up EFT}},  {\em Eur. Phys. J.
  C} {\bf 84} (2024), no.~5 458, [\href{http://arxiv.org/abs/2401.06214}{{\tt
  arXiv:2401.06214}}].

\bibitem{SINDRUMII:2006dvw}
{\bf SINDRUM II} Collaboration, W.~H. Bertl et~al., {\it {A Search for muon to
  electron conversion in muonic gold}},  {\em Eur. Phys. J. C} {\bf 47} (2006)
  337--346.

\end{thebibliography}\endgroup

\end{document}